\documentclass[twocolumn]{article}

\usepackage{geometry}
\usepackage{graphicx}

\usepackage{mathptmx}      

\newcommand{ \V }[1]{ \underline{#1} }
\newcommand{ \M }[1]{ \underline{\underline{#1}} }
\RequirePackage{amsmath,amssymb,bm}
\usepackage[ruled,section]{algorithm}
\usepackage{algorithmic,url,rotating}
\usepackage{subfigure}

\newcommand{\proj}{\ensuremath{\mathbf{P}}}

\title{A three-scale domain decomposition method for the 3D analysis of debonding in laminates. }

\author{P.~Kerfriden, O.~Allix, P.~Gosselet \vspace{5pt} \\ 
\large \textit{LMT-Cachan (ENS Cachan/CNRS/UPMC/PRES UniverSud Paris),} \\
\textit{61 av. du Pr\'esident Wilson, F-94230 Cachan, France}}

\date{January, 2009}

\begin{document}

\maketitle

The prediction of the quasi-static response of industrial laminate structures requires to use fine descriptions of the material, especially when debonding is involved. Even when modeled at the mesoscale, the computation of these structures results in very large numerical problems. In this paper, the exact mesoscale solution is sought using parallel iterative solvers. The LaTIn-based mixed domain decomposition method makes it very easy to handle the complex description of the structure; moreover the provided multiscale features enable us to deal with numerical difficulties at their natural scale; we present the various enhancements we developed to ensure the scalability of the method. An extension of the method designed to handle instabilities is also presented.

\section{Introduction}

Since the early 1980's, a very large number of studies has been conducted on the prediction of debonding in composite laminates, resulting in better understanding of the failure processes of composites. As a result, microme\-cha\-nics-based models have been shown to enable accurate predictions of the de\-bon\-ding phenomenon.

The industrialists' wish to replace expensive experiments by virtual tests for the design of their composite structures has brought about a new issue. Currently, the analysis of industrial problems using previously referenced micromodels is infeasible because the memory size and computing time requirements are far too large. Therefore, most applications to predict the initiation and propagation of debonding in laminates rely on mesoscale modeling \cite{deborst06,ladeveze02c} which are also rooted in the analysis of micromechanics phenomena. In this paper we retain the model described in \cite{allix92,allix98}; it describes the behavior of the laminates distinctly in the plies, 3D entities, and in the interfaces, 2D entities, the debonding ability being localized in the interfaces and handled through a cohesive behavior.

Even for small test cases, the numerical problem resulting from the mesomodeling of laminate structures is huge. Nevertheless, the latest advances in domain decomposition and multiscale methods provide powerful tools enabling the calculation of laminate structures of reasonable size. We can distinguish two families of solvers able to handle such large numerical problems. The first one consists in using a nonlinear homogenized strategy \cite{sanchezpalencia80,fish97,feyel00} coupled with local enrichment methods \cite{melenk96,hughes98,oden99,moes99,ghosh01,oliver04}. 
In this paper we focus on the prediction of debonding in process zones within the structure of the laminate, in which we consider that the debonding processes leading to final failure can be circumscribed. Though, within these potentially large process zones, no low gradient zone can be identified \textit{a priori}. Thus, the enrichment-based strategy might be difficult to use. Instead, we wish to use the second family of solvers for large numerical problems, which consists in computing the exact mesosolution everywhere in the process zones, using parallel iterative solvers \cite{farhat91,mandel93,ladeveze00,gosselet06,germain07b,dostal07}. Coupling the solution with a reduced model such as a plate model \cite{stein97,bendhia05} will be the subject of further work, and is not dealt with in this paper.


The mixed domain decomposition strategy described in \cite{ladeveze03b} uses an original concept which consists in splitting the structure in volume substructures separated by surface interfaces. Consequently, the reference problem resulting from the chosen mesomodeling is very naturally substructured, the cohesive interfaces of the model being handled within the interfaces of the domain decomposition method. This idea is developed in Section (\ref{sec:reference_problem}). Furthermore, the resolution of the substructured problem by a LaTIn iterative solver exhibits very interesting numerical properties: the nonlinearities are dealt with through local problems and very few re-assembling steps are required. The incremental micro-macro LaTIn algorithm is presented in Section (\ref{sec:multiscale_resolution}) without any improvement. As shown in Section (\ref{sec:delam_examp}) several numerical difficulties are encountered when directly applying the method. The core of the paper (Sections \ref{sec:ddr} to \ref{sec:riks}) presents and assesses improvements and adaptations of the method to efficiently handle delamination computations.

The substructured LaTIn method is parametrized by two interface search direction operators. Optimal values have been determined and practical values have been assessed for many mechanical problems (perfect interfaces, contact with or without friction), though for cohesive interfaces an adaptation is required in order the method to be efficient (and in some case feasible), as explained in Section (\ref{sec:ddr}).

The method is granted a multiscale nature by the separation of a macroscopic part and a microscopic part of the interface fields. This separation, coupled with continuity conditions, leads to the construction of an automatic homogenization procedure. This concept has been successfully applied in the cases of crack propagation problems in homogeneous media in \cite{guidault08}, of composite structures modeled at the microscale \cite{ladeveze06}, and of multiscale problems in time and space \cite{ladeveze08}. Though, in the case of singularities resulting from the crack tips being localized on the interfaces of the domain decomposition strategy, the homogeneous solution is too poor to represent accurately the solution. It results in a loss of extensibility of the strategy. Therefore, we enhance the method with a specific technique for the calculation of the quantities in the process zone with increased accuracy, which results in a significant improvement of the convergence rate, which is the topic of Section (\ref{sec:subiterations}).

A consequence to the substructuring naturally introduced to solve the reference problem is that the number of substructures required to solve large delamination problems becomes huge. Hence, the macroscopic problem can not be solved using direct solvers. The introduction of a third-level problem is required to quickly propagate the ve\-ry-high-wa\-ve\-length part of the solution. This is achieved by solving the second-scale problem using the balancing domain decomposition method described in \cite{mandel93}, as explained in Section (\ref{sec:third_scale}).

When trying to predict the very final residual strength of the structure, which is of the industrialist's interest, one has to deal with instabilities and limit-points problems resulting from the local softening behavior. Hence, an adaptation of the three-scale domain decomposition method to arc-length-type algorithms with local control of the loading amplitude has been developed, which is the subject of Section (\ref{sec:riks}).

\section{The reference problem and strategy}
\label{sec:reference_problem}
\subsection{Reference problem and substructuring}


\begin{figure}[h]
       \centering
       \includegraphics[width=0.99 \linewidth]{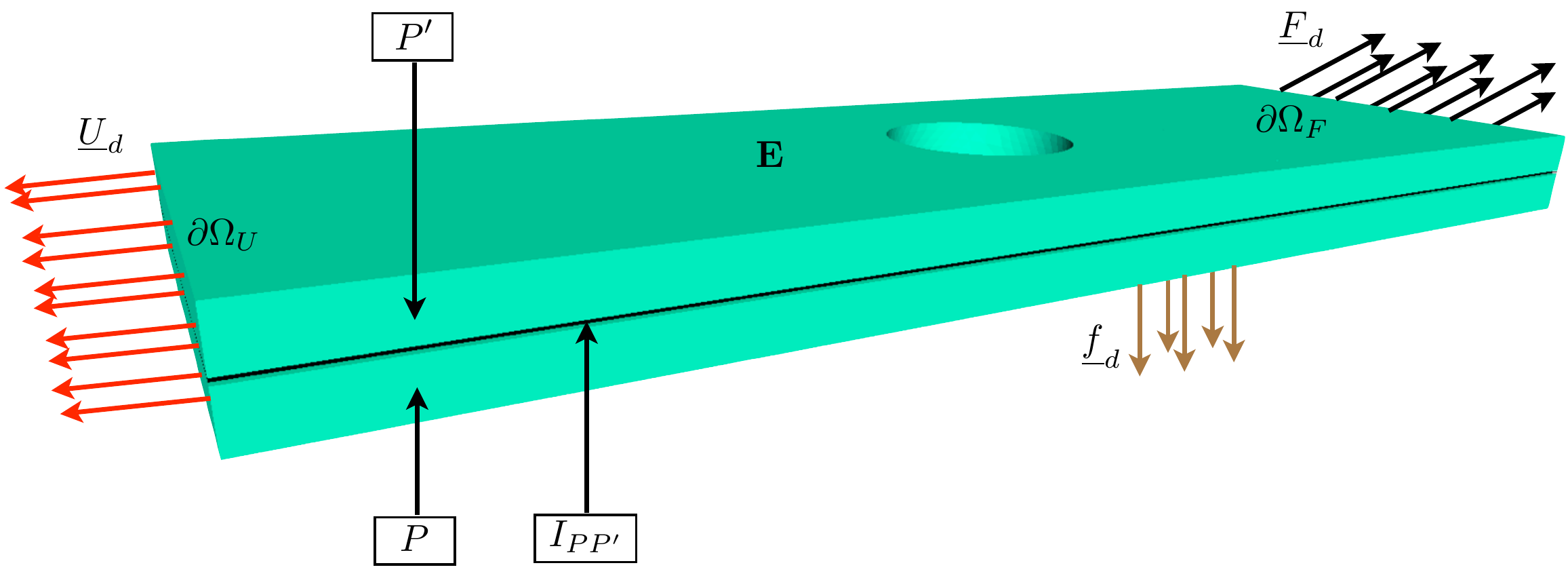}
       \caption{Reference problem}
       \label{fig:notations}
\end{figure}

The laminate structure $\mathbf{E}$ occupying the domain $\Omega$ is made out of $N_P$ adjacent plies $P$ occupying Domain $\Omega_P$, separated by $N_P-1$ cohesive interfaces $I_{PP'}$ (see Figure (\ref{fig:notations})).  An external traction field $\V{F}_d$ (respectively a displacement field $\V{U}_d$) is applied to the structure on Part $\partial \Omega_f$ (respectively $ \partial \Omega_u$) of the boundary $\partial \Omega$ of Domain $\Omega = \bigcup_P \Omega_P$. The normal to the boundary $\partial \Omega_P$ of Ply P, external to $P$, is $\V{n}_P$. The volume force is denoted $\V{f}_d$.
Let $\V{u}_P$ be the displacement field, $\M{\sigma}_P$  the Cauchy stress tensor and $\M{\epsilon}_P$ the symmetric part of the displacement gradient in Ply $P$.

The simulation is performed under the assumption of small perturbations and the evolution over time is
considered to be quasi-static and isothermal. The problem is solved using an implicit time integration scheme.  At each time step of the analysis, the reference equilibrium problem is: 

\vspace{5pt}
\textit{Find $\displaystyle s_{ref} = (s_P)_{P \in \mathbf{E}} $, where $\displaystyle s_P= (\M{\sigma}_P,\V{u}_P)$, which verifies the following equations:}
\begin{itemize}
\item Kinematic admissibility on $\partial \Omega_P \cap \partial \Omega_u$:
\begin{equation}
\V{u}_P = \V{U}_d
\end{equation}
\item Global equilibrium of Structure $\mathbf{E}$:
\begin{equation}
\begin{array}{l}
\displaystyle \forall \displaystyle {({\V{u}_P}^{\star})}_{P \in \mathbf{E}} \in {{\mathcal{U}_1}^0} \times ... \times {{\mathcal{U}_{N_P}}^0},
\\
\begin{array}{ll}
\displaystyle \sum_P &  \displaystyle \int_{\Omega_P}   \operatorname{Tr} \left( \M{\sigma}_P \M{\epsilon}({\V{u}_P}^{\star}) \right) d \Omega   
\\
& \displaystyle - \sum_P \int_{\Omega_P} \V{f}_d . {\V{u}_P}^{\star} \ d\Omega - \sum_P \int_{\partial \Omega \cap \partial \Omega_f} \V{F}_d . {\V{u}_P}^{\star} \ d\Gamma
\\
& \displaystyle + \sum_{P} \sum_{P' >P} \int_{\partial \Omega_P \cap \partial \Omega_{P'}} \M{\sigma}_P \V{n}_p . {\V{[ u ]}_P}^{\star} \ d\Gamma = 0
\end{array}
\end{array}
\end{equation}
\item Linear orthotropic behavior of the plies:
\begin{equation}
\textrm{at each point of } \Omega_P, \quad \M{\sigma}_P = K \, \M{\epsilon}(\underline{u}_P)
\end{equation}
\item Constitutive equation of the interfaces:
\begin{equation}
\begin{array}{l}
\displaystyle \textrm{at each point of } I_{PP'}, \quad
\displaystyle {\mathcal{A}}_{PP'} (   \,\V{[u]}_P \, , \, \M{\sigma}_P \V{n}_P \, ) = 0
\end{array}
\end{equation}
\end{itemize}

\begin{figure}[htb]
       \centering
       \includegraphics[width=0.99 \linewidth]{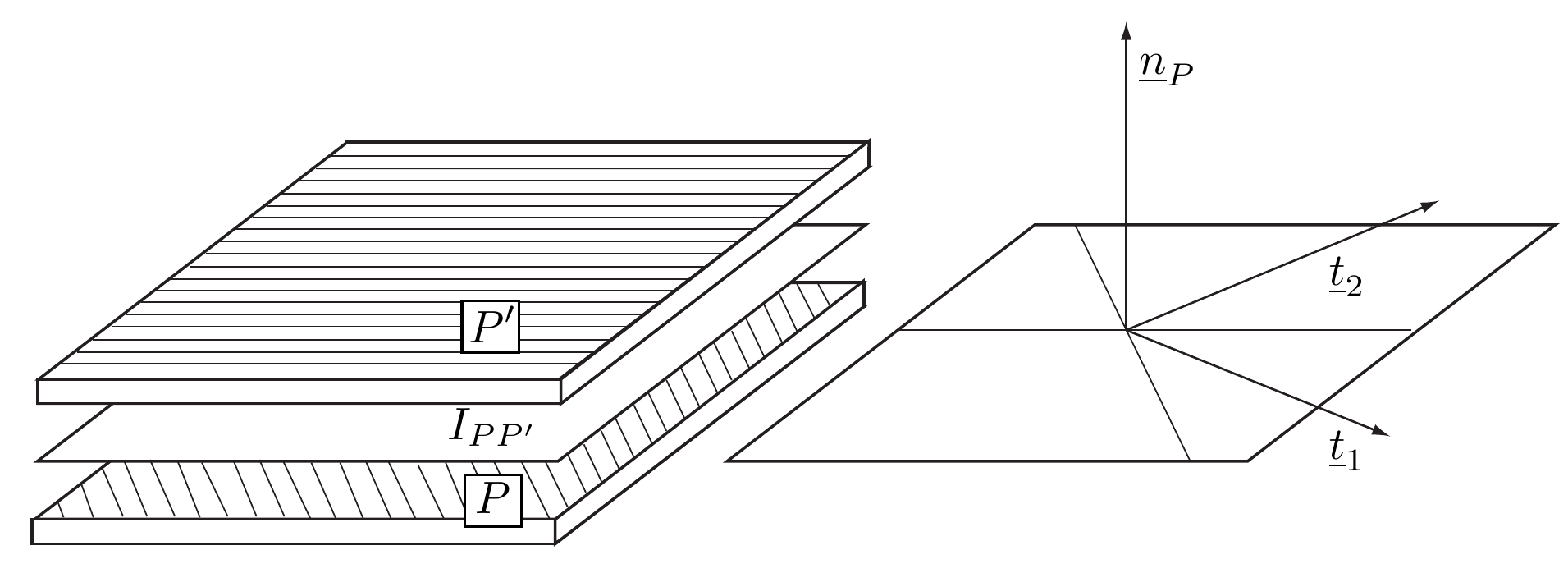}
       \caption{The mesomodel entities}
       \label{fig:interface_composite}
\end{figure}

The gap of displacement $\V{[u]}_P$ of Interface $I_{PP'}$ such that $P<P'$ has arbitrarily been introduced as $\V{[u]}_P = \V{u}_{P'}-\V{u}_{P}$.
The operator $\mathcal{A}_{PP'}$ establishes a relation between the primal interface unknown $\V{[u]}_P$, and the dual interface unknown $\M{\sigma}_P \V{n}_P$, which reads :  
\begin{equation}
\displaystyle \M{\sigma}_P.\V{n}_P = K_{PP'}. \V{[u]}_P \\
\end{equation}
The expression of the local stiffness operator $\M{K}_{PP'}$ of Interface $I_{PP'}$ can be made explicit in the basis $(\V{n}_P,\V{t}_1,\V{t}_2,)$ (see Figure (\ref{fig:interface_composite})):
\begin{equation}
\nonumber
\M{K}_{PP'} = \left( \begin{array}{ccc}
\displaystyle \left(1-h_+(\V{[u]}_P.\V{n}_P)\ d_3 \right) k_n^0  & 0 & 0 \\
0 & \displaystyle (1-d_1)\, k_t^0 & 0 \\
0 & 0 & \displaystyle (1-d_2)\, k_t^0 
\end{array}
\right)
\end{equation}
$h_+$ is here the positive indicator function.

The local damage variables ${d_i}$ are introduced into the interface model in
order to simulate its softening behavior when the structure is
loaded. Their values range from $0$ (healthy
interface point) to $1$ (completely damaged interface point).
The parameters $d_i$ are related to the local energy release rates
$Y_i$ of the interface's degradation modes. Denoting $e_d$ the surface strain energy of the cohesive interface,
\begin{equation}
Y_i = - \frac{\partial e_d}{\partial d_i}
\qquad \textrm{where} \quad
\left\{ \begin{array}{ccl}
Y_1 & = & \displaystyle \frac{1}{2} \, k_t^0 \, (\V{[u]}_P.\V{t}_1)^2 \\
Y_2 & = & \displaystyle \frac{1}{2} \, k_t^0 \, (\V{[u]}_P.\V{t}_2])^2 \\
Y_3 & = & \displaystyle \frac{1}{2} \, k_n^0 \, \left( h_+(\V{[u]}_{P}.\V{n}_P ) \right)^2
\end{array} \right. \end{equation}
$e_d$ is here the surface strain energy of the cohesive interface.

We assume that the damage variables are functions of a
single quantity: the maximum $Y_{|t}$ over time of a combination of
the energy release rates ${Y_i}_{|\tau}$, $\tau \leq t$:
\begin{equation}
Y_{|t} =  \operatorname{max}_{(\tau \leq t)} \left( {Y_3}_{|\tau}^\alpha + \gamma_1 {Y_1}_{|\tau}^\alpha + \gamma_2 {Y_2}_{|\tau}^\alpha \right)^{\frac{1}{\alpha}}
\end{equation}
Thus, the evolution laws are:
\begin{equation}
d_1 = d_2 = d_3 = w(Y) 
\end{equation}
\begin{equation}
\nonumber
\textrm{where in general} \quad w(Y) = \frac{n}{n+1} \left( \frac{Y}{Y_c} \right)^n
\end{equation}

This model has the advantage of using a single damage variable to
handle several macroscopic delamination modes of the
interface (traction along $\V{n}_P$ and shear along $\V{t}_1$ and $\V{t}_2$). However,
when setting Parameters $\gamma_1$ and $\gamma_2$ to identified
physical values such that $\gamma_1 \neq \gamma_2 \neq 1$, the
energy dissipated due to the propagation of the crack is different
for the three modes.

\begin{figure*}
       \centering
       \includegraphics[width=0.8 \linewidth]{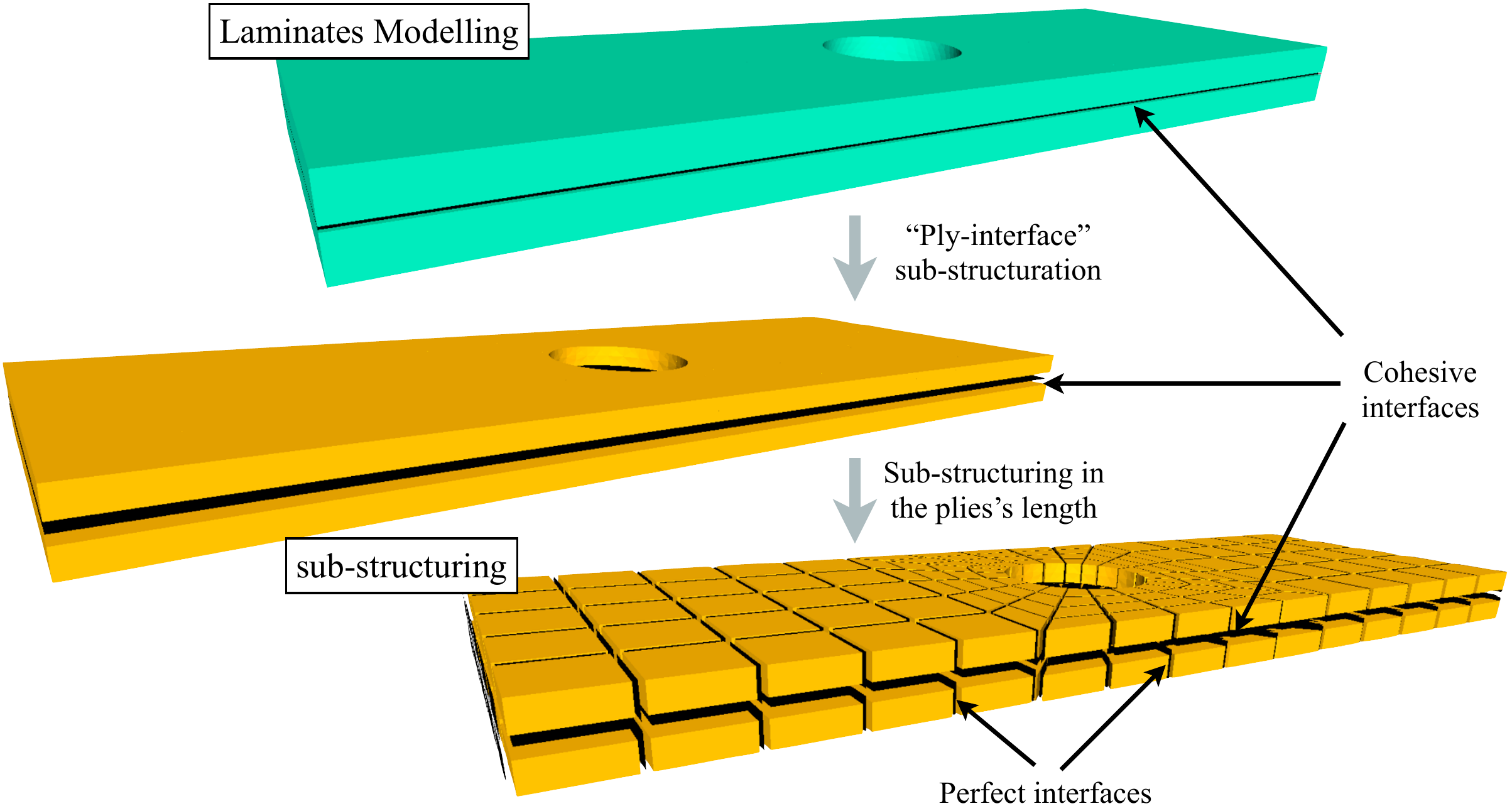}
       \caption{Substructuring of the laminated composite structure}
       \label{fig:decomp_sst_interfaces}
\end{figure*}

\subsubsection{Substructured formulation of the reference problem}

\begin{figure}
       \centering
       \includegraphics[width=0.95 \linewidth]{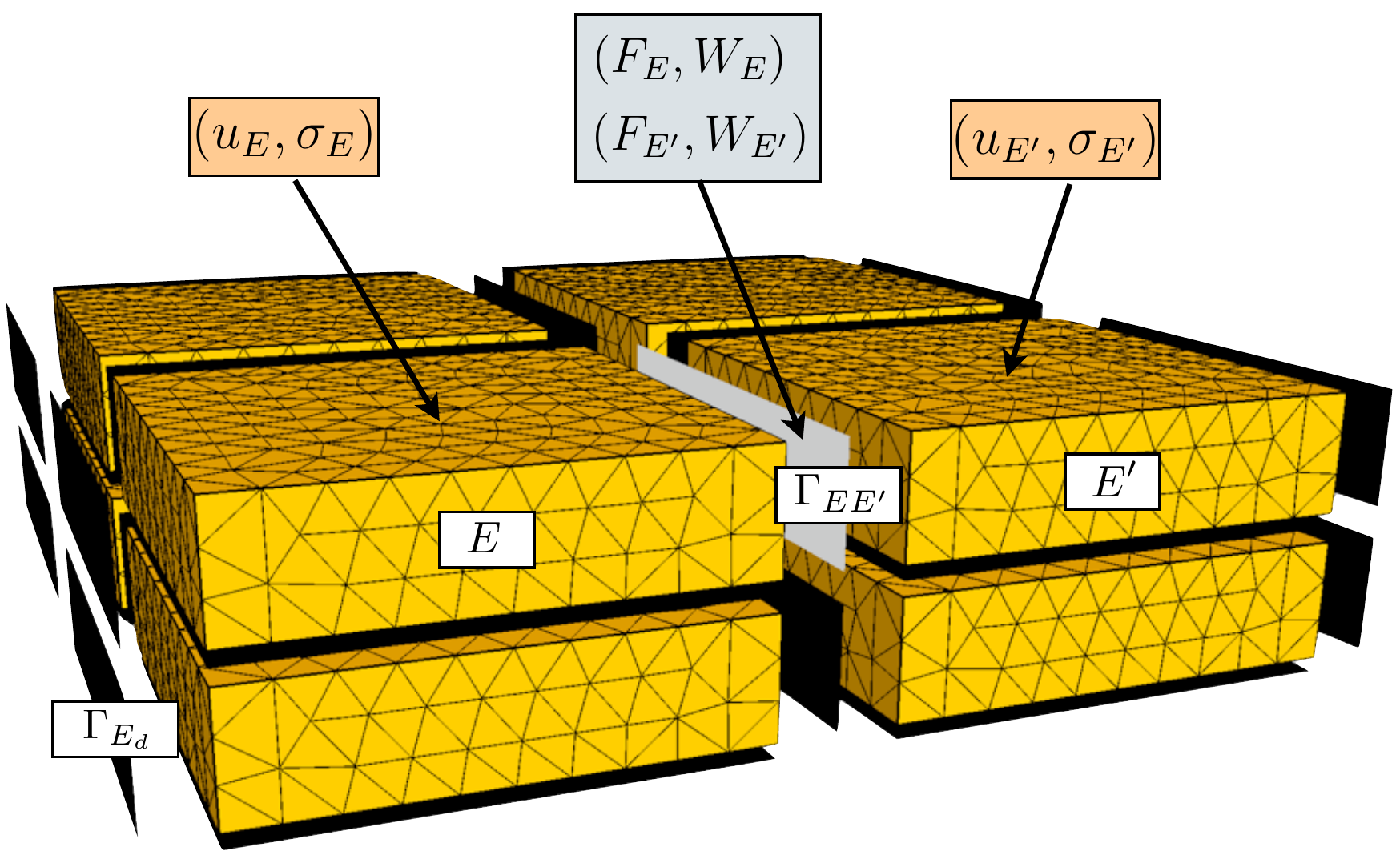}
       \caption{Substructuring of the laminated composite structure}
       \label{fig:champs_interface}
\end{figure}

The laminates structure $\mathbf{E}$ is decomposed into substructures and interfaces as represented in Figure (\ref{fig:decomp_sst_interfaces}). Each of these mechanical entities possesses its own kinematic and static unknown fields linked by its behavior. The substructuring is driven by the will to match domain decomposition interfaces with material cohesive interfaces, so that each substructure belongs to a unique ply $P$ and has a constant linear behavior. A substructure $E$ defined in Domain $\Omega_E$ is connected to an adjacent substructure $E'$ through an interface $\Gamma_{EE'}=\partial \Omega_E \cap\partial \Omega_{E'}$ (Figure (\ref{fig:champs_interface})). The surface entity $\Gamma_{EE'}$ applies force distributions $\underline{F}_E$, $\underline{F}_{E'}$ as well as displacement distributions $\underline{W}_E$, $\underline{W}_{E'}$ to $E$ and $E'$ respectively. We write $\Gamma_E = \bigcup_{E' \in \mathbf{E}} \Gamma_{EE'} $.


On a substructure $E$ such that $\Gamma_E \cap ( \partial \Omega_f
\cup \partial \Omega_u) \neq \{0\}$, the boundary condition $(\V{U}_d,\V{F}_d)$ is
applied through a boundary interface $\Gamma_{E_d}$.

Let $\M{\sigma}_E$ be the Cauchy stress tensor and $\M{\epsilon}_E$ the
symmetric part of the displacement gradient in substructure $E$.

Then, the substructured quasi-static problem consists in finding $s=
(s_E)_{E \in \mathbf{E}} $ at a given step of the time integration scheme, where $s_E = (\V{u}_E , \V{W}_E ,
\M{\sigma}_E , \V{F}_E )$, which verifies the following equations:
\begin{itemize}
\item Kinematic admissibility of Substructure $E$:
\begin{equation}
\textrm{at each point of } \Gamma_E, \quad {\V{u}_E} = {\V{W}_{E}}
\end{equation}
\item Static admissibility of Substructure $E$:
\begin{equation}
\label{equation:equilibre_ss}
 \begin{array}{l}
\displaystyle \forall ({\underline{u}_E}^\star,{\underline{W}_E}^\star) \in \mathcal{U}_{E} \times \mathcal{W}_{E} \  / \  {{\V{u}_E}^\star}_{| \partial \Omega_E} = {{\V{W}_{E}}^\star},  \\
\begin{array}{ll}
\displaystyle \int_{\Omega_E} \operatorname{Tr} \left( \M{\sigma}_E \, \M{\epsilon} ({\V{u}_E}^\star) \right) \ d \Omega & = \displaystyle \int_{\Omega_E} \V{f}_d . {\V{u}_E}^\star \, d \Omega \\
&+\displaystyle  \int_{\Gamma_E} \V{F}_E . {\V{W}_E}^\star \ d \Gamma  
\end{array}
\end{array}
\end{equation}
\item Linear orthotropic behavior of Substructure $E$:
\begin{equation}
\textrm{at each point of } \Omega_E,\quad \M{\sigma}_E = K \, \M{\epsilon}(\underline{u}_E)
\end{equation}
\item Behavior of the interfaces $\Gamma_{EE'}$:
\begin{equation}
\begin{array}{l}
\displaystyle \textrm{at each point of } \Gamma_{EE'} \in \Gamma_E, \\ 
\displaystyle \mathcal{R}_{EE'}( \V{W}_{E}  , \V{W}_{E'} , \V{F}_{E} , \V{F}_{E'} ) = 0
\end{array}
\end{equation}
\item Behavior of the interface at the boundary:
\begin{equation}
\textrm{at each point of } \Gamma_{{E}_d}, \quad \mathcal{R}_{E_d}( \V{W}_{E}, \V{F}_{E}) = 0
\end{equation}
\end{itemize}

The formal relation $\mathcal{R}_{EE'}=0$ and $\mathcal{R}_{E_d}=0$ named "interface behavior" can be made explicit in the two cases that we have to handle:
\begin{itemize}
\item Perfect interface:
\begin{equation}
\left\{ \begin{array}{l}
\V{F}_{E} + \V{F}_{E'} = 0 
\\
\V{W}_{E} - \V{W}_{E'} = 0
\end{array} \right.
\end{equation}
\item Cohesive interface:
\begin{equation}
\left\{ \begin{array}{l}
\V{F}_{E} + \V{F}_{E'} = 0
\\
\displaystyle \mathcal{A}_{PP'}(\, \V{W}_{E'} - \V{W}_{E}  \, ,\, \V{F}_{E} \, ) = 0
\end{array} \right.
\end{equation}
where Substructure $E$ (respectively $E'$) belongs to Ply $P$ (respectively $P'$), such that $P<P'$.
\end{itemize}

In the same manner, and for instance in the case of a prescribed displacement boudary interface $\Gamma_{E_d}$, the formal relation $\mathcal{R}_{E_d}=0$ reads:
\begin{equation}
\V{W}_{E} = \V{U}_{d}
\end{equation}

\subsection{Two-scale iterative resolution of the substructured problem}
\label{sec:multiscale_resolution}

\subsubsection{Introduction of the macroscopic scale}

In order the scalability of the method to be ensured, a global coarse grid problem is solved at each iteration of the solver. The definition of the macroscopic fields required to construct this problem is done on the interface only.



\begin{figure*}
       \centering
       \includegraphics[width=0.8 \linewidth]{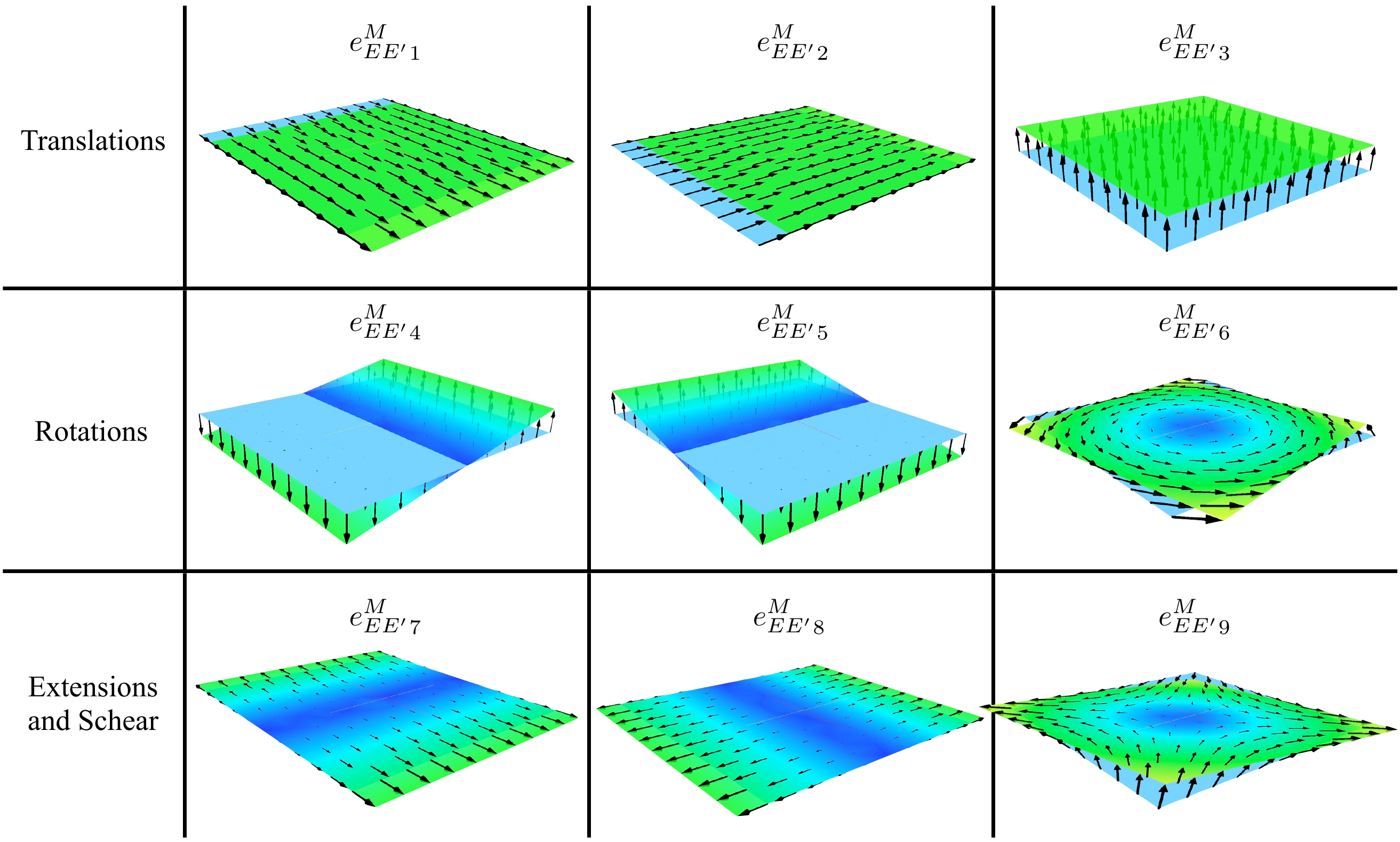}
       \caption{The linear macro basis for a plane interface}
       \label{fig:base_macro}
\end{figure*}

At each interface $\Gamma_{EE'}$, the interface fields are split
into a macro part $^M$ and a micro part $^m$, the former belonging
to a small-dimension subspace (e.g. 4 macro degrees of freedom per
interface in 2D, 9 in 3D).
\begin{equation}
\begin{array}{l}
\displaystyle {\underline{F}_E} = \underline{F}_E^M + \underline{F}_E^m \\
\displaystyle {\underline{W}_E} = \underline{W}_E^M + \underline{W}_E^m
\end{array}
\end{equation}
Given the macrospaces $\mathcal{W}_E^M$ and $\mathcal{F}_E^M$ on Interface $\Gamma_E$, the unicity of the decomposition of the interface fields in macro and micro data is ensured by uncoupling the interface virtual works:
\begin{equation}
\begin{array}{ll}
\displaystyle \forall  (\underline{F}_E,\underline{W}_E)  \in & \displaystyle \mathcal{F}_E\times\mathcal{W}_E, \quad 
 \int_{\Gamma_{EE'}} \underline{F}_E.\underline{W}_E \ d \Gamma 
\\ 
&   \displaystyle 
=  \int_{\Gamma_{EE'}} \underline{F}_E^M.\underline{W}_E^M \ d \Gamma  
  + \int_{\Gamma_{EE'}} \underline{F}_E^m.\underline{W}_E^m \ d \Gamma
\end{array}
\end{equation}
Usually, a common macro basis for both the kinematic and static interface macro fields is chosen. Numerical tests showed that in order to ensure the numerical scalability of the method the macro basis should extract at least the linear part of the interface forces (see Figures (\ref{fig:base_macro}) and
(\ref{fig:DCB})). Indeed, this macro space contains the part of the interface fields with the highest wavelength. Consequently, according to the Saint-Venant principle, the micro complement (which, as explained in next subsection, is found iteratively through the resolution of local problems) has only local influence.

\subsubsection{The iterative algorithm}

The iterative LaTIn algorithm, which enables the resolution of nonlinear problems, is here applied to the resolution of the substructured reference problem with nonlinearities localized in the interfaces. 

The equations of the problem are split into two groups:
\begin{itemize}
    \item linear equations in substructure and macroscopic interface variables:
    \begin{itemize}
        \item[$\bullet$] static admissibility of the substructures
        \item[$\bullet$] kinematic admissibility of the substructures
        \item[$\bullet$] linear behavior of the substructures
        \item[$\bullet$] equilibrium of the macro interface forces
    \end{itemize}
    \item local equations in interface variables:
    \begin{itemize}
        \item[$\bullet$] interface behavior
    \end{itemize}
\end{itemize}

The interface solutions 
$\displaystyle \mathbf{s} = (\mathbf{s}_E)_{E \in \mathbf{E}} = (\V{W}_E , \V{F}_E)_{E \in \mathbf{E}}$ 
to the first set of equations belong to Space $\mathbf{A_d}$, while the interface solutions
$\displaystyle \mathbf{\widehat{s}} = (\mathbf{\widehat{s}}_E)_{E \in \mathbf{E}} = (\V{\widehat{W}}_E , \V{\widehat{F}}_E )_{E \in \mathbf{E}}$ 
to the second set of equations belong to $\boldsymbol{\Gamma}$. The converged interface solution $\mathbf{s}_{\textbf{ref}}$ is such that:
\begin{equation}
\mathbf{s}_{\textbf{ref}} \in \mathbf{A_d} \bigcap \boldsymbol{\Gamma}
\end{equation}

\begin{figure}
       \centering
       \includegraphics[width=0.75 \linewidth]{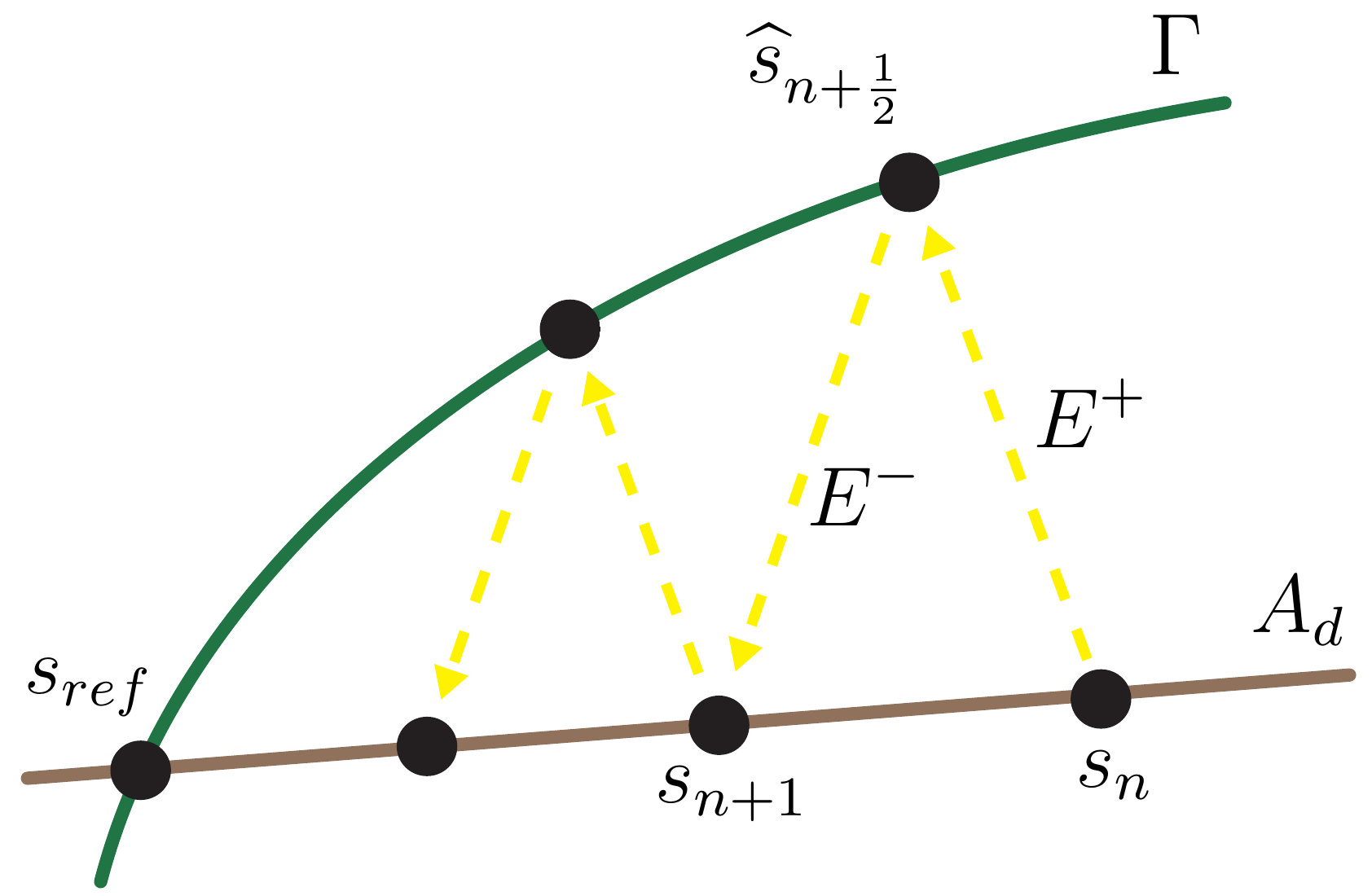}
       \caption{Substructuring of the laminated composite structure}
       \label{fig:latin}
\end{figure}

The resolution scheme consists in seeking the interface solution $\mathbf{s}_{\textbf{ref}}$
alternatively in these two spaces: first, one finds a solution $\mathbf{s}_n$
in $\mathbf{A_d}$, then a solution $\mathbf{\widehat{s}}_{n+\frac{1}{2}}$ in
$\boldsymbol{\Gamma}$. In order for the two problems to be
well-posed, search directions $\mathbf{E}^+$ and $\mathbf{E}^-$
linking the solutions $\mathbf{s}$ and $\mathbf{\widehat{s}} \,$ through the iterative
process are introduced (see Figure (\ref{fig:latin})).

Hence, an iteration of the resolution scheme consists of two stages:
\begin{itemize}
\item a local stage:
\begin{equation}
\textit{Find} \ \mathbf{\widehat{s}}_{n+\frac{1}{2}} \in \boldsymbol{\Gamma} \ \textit{such that}  \  \left( \mathbf{\widehat{s}}_{n+\frac{1}{2}} - \mathbf{s}_{n} \right) \in \mathbf{E}^+
\end{equation}
\item a linear stage:
\begin{equation}
\textit{Find} \ \mathbf{s}_{n+1} \in \mathbf{A_d} \ \textit{such that}  \  \left( \mathbf{s}_{n+1} - \mathbf{\widehat{s}}_{n+\frac{1}{2}} \right) \in \mathbf{E}^-
\end{equation}
\end{itemize}
In the following sections, the subscript $n$ will be dropped.

\paragraph{Local stage}

In the local stage, local problems are solved at each point of the interfaces $\Gamma_{EE'}$:
\begin{equation}
\label{eq:local_problem}
\begin{array}{l}
\textit{Find} \ (\underline{\widehat{F}}_E,\underline{\widehat{W}}_E,\underline{\widehat{F}}_{E'},\underline{\widehat{W}}_{E'})  \vspace{0.2cm}
\textit{ such that: } \\
\left\{ \begin{array}{l}
\displaystyle \mathcal{R}_{EE'}(\underline{\widehat{W}}_E,\underline{\widehat{W}}_{E'},\underline{\widehat{F}}_E,\underline{\widehat{F}}_{E'}) = 0  \\
\displaystyle  (\underline{\widehat{F}}_E-\underline{F}_E) - k_E^+ \, (\underline{\widehat{W}}_E - \underline{W}_E) = 0 \\
\displaystyle  (\underline{\widehat{F}}_{E'}-\underline{F}_{E'}) - k_{E'}^+ \, (\underline{\widehat{W}}_{E'} - \underline{W}_{E'}) = 0
\end{array} \right.
\end{array}
\end{equation}
the two last equations of this system being the search direction $\mathbf{E}^+$. In the case of a cohesive interface, Problem \eqref{eq:local_problem} is nonlinear, and solved by a modified Newton-Raphson scheme.

Local linear problems are also solved at each point of the boundary interfaces $\Gamma_{E_d}$:
\begin{equation}
\begin{array}{l}
\textit{Find} \ (\underline{\widehat{F}}_E,\underline{\widehat{W}}_E) \textit{such that: }
\\
\left\{ \begin{array}{l}
\displaystyle \mathcal{R}_{E_d}(\underline{\widehat{W}}_E,\underline{\widehat{F}}_E) = 0 \\
\displaystyle  (\underline{\widehat{F}}_E-\underline{F}_E) - k_E^+ (\underline{\widehat{W}}_E - \underline{W}_E) = 0 
\end{array} \right.
\end{array}
\end{equation}

\paragraph{Linear stage}



The linear stage consists in solving linear systems on each substructure under the constraint of macroscopic equilibrium of the interface forces.
\begin{equation}
\label{eq:macro_ad}
\text{on interface }\Gamma_{EE'},\quad \underline{F}_E^M + \underline{F}_{E'}^M =  0
\end{equation}
Macroscopic admissibility of displacements could also be enforced. In the case of 
perfect interfaces, it would be easy to derive. In the case of non-homogeneous or nonlinear behavior at the
interfaces, the macro condition would not be practically (and sometimes theoretically) feasible. 

Condition \eqref{eq:macro_ad} is incompatible with the monoscale search direction $\mathbf{E}^-$ coupling the interface displacement and forces fields at the linear stage, which reads:
\begin{equation}
\begin{array}{l}
\displaystyle \text{on interface }\Gamma_E,  \quad
\displaystyle (\underline{F}_{E}-\underline{\widehat{F}}_{E}) + k_E^- \ (\underline{W}_{E}-\underline{\widehat{W}}_{E}) = 0
\end{array}
\end{equation}
Hence this search direction is weakened and verified at best under the macroscopic constraint. Technically this is realized using a Lagrangian whose
stationarity leads to a modified local search direction:
\begin{equation} \label{eq:ddr_loc}
\begin{array}{ll}
\displaystyle \forall & \displaystyle {\underline{W}_E}^\star \in \mathcal{W}_E, \quad \int_{\Gamma_E}  (\underline{F}_{E}-   \underline{\widehat{F}}_{E}) \, . \, {\underline{W}_E}^\star \ d\Gamma
 \\
& \displaystyle + \int_{\Gamma_E}  \left( k_E^- \ (\underline{W}_{E}-\underline{\widehat{W}}_{E})   -   k_E^- \, \underline{\widetilde{W}}^M \right) \, . \, {\underline{W}_E}^\star \ d\Gamma= 0
\end{array}
\end{equation}
and to an equation expressing the continuity of the macroforces:
\begin{equation}\label{eq:ddr_macro}
\begin{array}{l}
\displaystyle \forall \ \underline {\widetilde{W}}^{M\star} \in {\mathcal{W}^{M}},   \\
 \displaystyle \quad \sum_E \int_{\Gamma_{E}} \underline{F}_{E} . \underline {\widetilde{W}}^{M\star} \ d\Gamma
 = 
 \sum_E \int_{\partial \Omega_f} \underline{F}_d . \underline{\widetilde{W}}^{M\star} \ d\Gamma
 \end{array}
\end{equation}
where $\underline{\widetilde{W}}^{M}$ is unique for Interface $\Gamma_{EE'}$ and set to zero on $\partial_u\Omega$.

A way to solve this set of equations consists in introducing a
relation coupling $ \underline{F}_{E}^M$ and $\underline
{\widetilde{W}}^M$ into \eqref{eq:ddr_macro}. This relation is
derived from the local equilibrium of each substructure \eqref{equation:equilibre_ss}
and from the local modified search direction
\eqref{eq:ddr_loc}. The problem to be solved for Substructure $E$
becomes:
\begin{equation}
\label{eq:pb_micro}
\begin{array}{l}
\displaystyle \forall \ ({\V{u}_E}^\star,{\V{W}_E}^\star) \in {\mathcal{U}_E} \times {\mathcal{W}_E}, 
\\
\displaystyle  \int_{\Omega_E} \operatorname{Tr} (\M{\epsilon} (\V{u}_E) \,  K \M{\epsilon} ({\V{u}_E}^\star) ) \ d\Omega + \int_{\Gamma_E} k_E^- \, \V{W}_E . {\V{W}_E}^\star \ d \Gamma 
 \\
\displaystyle \quad =  \int_{\Omega_E} \V{f}_d . {\V{u}_E}^\star \, d \Omega + \int_{\Gamma_E} (\underline{\widehat{\widehat{F}}}_E +k_E^- \, \V{\widetilde{W}}^M) . {\V{W}_E}^\star \ d \Gamma
\end{array}
\end{equation}
where $\underline{\widehat{\widehat{F}}}_E  =
\underline{\widehat{F}}_E +k_E^- \, \underline{\widehat{W}}_E$.

Equation \eqref{eq:pb_micro} is linear. Therefore, one can write a
linear relation between the interface displacements and the loading:
\begin{equation}
\begin{array}{ll}
\displaystyle \forall \  {\underline{W}_E}^\star \in & \displaystyle \mathcal{W}_E ,     \quad \int_{\Gamma_E}  \underline{W}_E.{\underline{W}_E}^\star \ d\Gamma  \\ 
& \displaystyle = \int_{\Gamma_E} \left( \mathbb{H}_E (\underline{\widehat{\widehat{F}}}_E + k_E^-  \ \underline{\widetilde{W}}^M ) + {\V{W}_{E}^c}_d \right) .  {\underline{W}_E}^\star
\end{array}
\end{equation}
Operator $\mathbb{H}_E$ is the dual Schur complement of Substructure
$E$ modified by the search direction, while ${\V{W}_{E}^c}_d$ results from the condensation of the volumic loading on interface $\Gamma_E$.

The corresponding interface forces are obtained through the modified
search direction \eqref{eq:ddr_loc} and projected onto the macro
space:
\begin{equation}
\label{eq:comp_homo}
\begin{array}{l}
\displaystyle \forall \ \underline{\widetilde{W}}^{M \star} \in {\mathcal{W}^M} , \\
\displaystyle \quad 
\int_{\Gamma_E} \underline{F}_E . \underline{\widetilde{W}}^{M \star} \ d\Gamma 
\displaystyle = \int_{\Gamma_E} ( \mathbb{L}_E^M \ \underline{\widetilde{W}}^M + \underline{\widetilde{F}}_E^M) \, . \,  \underline{\widetilde{W}}^{M \star} \ d\Gamma
\end{array}
\end{equation}
where
\begin{equation}
\left\{ \begin{array}{ll}
\displaystyle \forall & \displaystyle \underline{\widetilde{W}}^{M \star} \in {\mathcal{W}^M}, \quad
\int_{\Gamma_E} \mathbb{L}_E^M \ \underline{\widetilde{W}}^M \, . \,  \underline{\widetilde{W}}^{M \star} \ d \Gamma 
\\ \displaystyle
& \displaystyle \qquad \qquad = \int_{\Gamma_E}  ( k_E^- - k_E^- \, \mathbb{H}_E \ k_E^- ) \ \underline{\widetilde{W}}^M  \, . \, \underline{\widetilde{W}}^{M \star} \ d\Gamma
\\
\displaystyle \forall & \displaystyle \underline{\widetilde{W}}^{M \star} \in {\mathcal{W}^M}, \quad
\int_{\Gamma_E} \underline{\widetilde{F}}_E^M \, . \, \underline{\widetilde{W}}^{M \star} \ d \Gamma 
\\ \displaystyle
& \displaystyle \qquad \qquad = \int_{\Gamma_E}  ( \widehat{\widehat{F}}_E
- k_E^- \ \left( \mathbb{H}_E \ \widehat{\widehat{F}}_E + {\V{W}_{E}^c}_d ) \right) \, . \, \underline{\widetilde{W}}^{M \star} \ d\Gamma
\nonumber
\end{array} \right.
\end{equation}
$\mathbb{L}_E^M$ is classically viewed as a homogenized behavior of
Substructure $E$ and is calculated explicitly for each substructure
by solving local subproblems \eqref{eq:pb_micro} taking the vectors
of the macro basis as boundary conditions on $\Gamma_E$.

This relation is finally introduced into the equation expressing the
admissibility of the macroforces \eqref{eq:ddr_macro}, leading to the so-called macro-problem:
\begin{equation}
\label{eq:pb_macro}
\begin{array}{ll}
\displaystyle \forall & \displaystyle \underline {\widetilde{W}}^{M \star} \in {\mathcal{W}^M}, \quad
\sum_E \int_{\Gamma_{E}} \mathbb{L}_E^M \ \underline{\widetilde{W}}^M . \underline {\widetilde{W}}^{M\star} \ d\Gamma
 \\ & \qquad \displaystyle =
 \sum_E \int_{\partial \Omega_f} \underline{F}_d .\underline{\widetilde{W}}^{M\star} \ d\Gamma
 -  \sum_E \int_{\Gamma_{E}}  \underline{\widetilde{F}}_E .\underline{\widetilde{W}}^{M\star} \ d\Gamma
\end{array}
\end{equation}
The macro-problem is discrete by nature. Hence, its an algebraic form $\displaystyle \mathbf{L^M} \ \widetilde{W}^M = F^M$, where $\widetilde{W}^M$ is the vector of the components of the Lagrange multiplier in the macro basis, is also used in the following.

The right-hand side of Equation \eqref{eq:pb_macro}  can be interpreted as a
macroscopic static residual obtained from the calculation of a
single-scale linear stage. In order to derive this term, the problem \eqref{eq:pb_micro} must be solved independently on each substructure, setting $\underline{\widetilde{W}}^M$ to zero. 
The resolution of the macroscopic problem  \eqref{eq:pb_macro} leads to the global knowledge of Lagrange multiplier $\underline{\widetilde{W}}^M$, which is finally used as prescribed displacement to solve the substructure independent problems \eqref{eq:pb_micro}.

In order to perform the resolutions of \eqref{eq:pb_micro} on the substructures, finite element method is used. Since the behavior of the substructures is linear, the stiffness operator of each substructure can be factorized once at the beginning of the calculation and reused without updating throughout the analysis, which gives the method high numerical performance.

Algorithm (\ref{alg:LaTIn}) sums up the iterative procedure which has been described in this section.

\begin{algorithm}[ht]\caption{The 2-scale domain decomposition solver}\label{alg:LaTIn}
\begin{algorithmic}[1]
\STATE Substructures' operators construction
\STATE Computation of the macroscopic homogenized behavior $\mathbb{L}_E^M$ on each substructure 
\STATE Global assembly of the macroscopic operator $\mathbf{L^M}$
\STATE Initialization $\mathbf{s}_0 \in \boldsymbol{\Gamma}$
\FOR{$n=0,\ldots,N$}%
  \STATE Linear stage: computation of $\mathbf{s}_n \in \mathbf{A_d}$ \\
   $\quad \square$ Computation of the macroscopic right-hand term $\widetilde{F}_E^M$ \\
   $\quad \ \  \ $ on each substructure \\
   $\quad \square$ Global assembly of the macroscopic right-hand term \\
   $\quad \square$ Macro problem resolution \\
   $\quad \square$ Micro problem resolution
  \STATE Local stage: computation of $\mathbf{\widehat{s}}_{n+\frac{1}{2}} \in \boldsymbol{\Gamma}$ \\
   $\quad \square$ Boundary interfaces $\Gamma_{E_d}$ \\
   $\quad \square$ Internal interfaces $\Gamma_{EE'}$
   \STATE Calculation of an error indicator
\ENDFOR
\end{algorithmic} \end{algorithm}

\subsection{Delamination analysis example}
\label{sec:delam_examp}

\begin{figure}
       \centering
       \includegraphics[width=0.99 \linewidth]{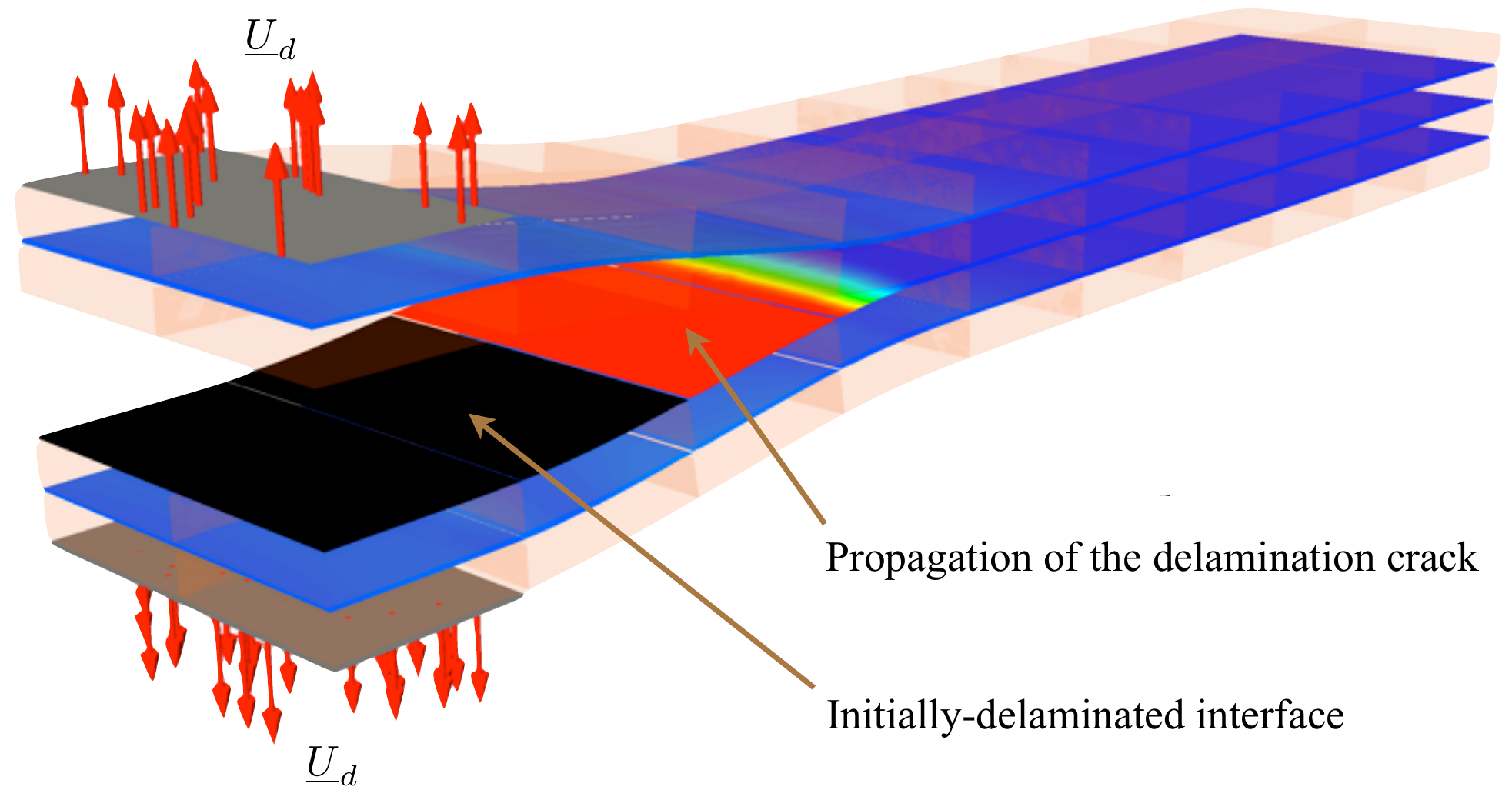}
       \caption{Four-ply DCB test case}
       \label{fig:DCB}
\end{figure}

A first example of quasi-static delamination analysis is shown in Figure (\ref{fig:DCB}). The problem consists in a $\textrm{[0/90]}_s$ double cantilever beam (DCB) case. The loading leading to mode I quasi-static crack's propagation is increased linearly over 10 time steps. The first three of them correspond to the initiation of the delamination and the remainder to the crack's propagation.

\vspace{5pt}
This assessment is realized with a C++ implementation of the mixed domain decomposition method capable of handling the quasi-static analysis of 3D nonlinear problems. The parallel computations use the MPI library to exchange data among several processors.

Each processor is assigned to a set of connected substructures (along with their interfaces); it calculates the associated operators and
solves the local problems. This tends to reduce the number of interfaces duplicated among several processors (Figure (\ref{fig:third_scale})) and is achieved technically through a METIS routine.

\begin{figure}
       \centering
       \includegraphics[width=0.99 \linewidth]{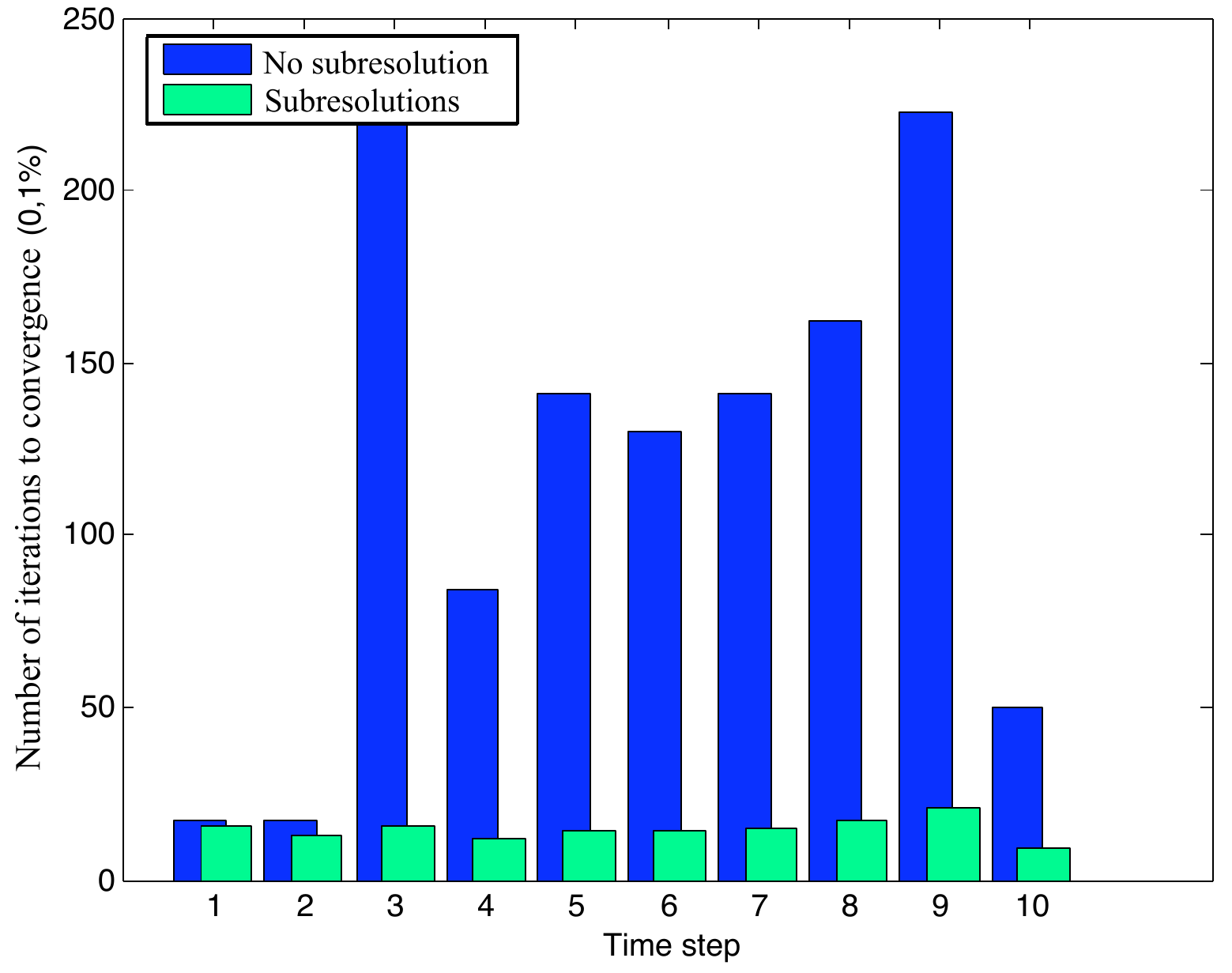}
       \caption{Subiterations around the crack's tip}
       \label{fig:sub_iterations}
\end{figure}

\vspace{5pt}
This very simple test case already exhibits numerical difficulties: 
\begin{itemize}
\item The convergence rate of the LaTIn-based strategy is highly dependant on the search direction parameters. In the case of cohesive interfaces the iterative solver can even stagnate when using too small values for the search direction parameters. Hence, we describe in next section the way we set them in order to ensure convergence.
\item The method loses its numerical scalability when the crack's tip propagates. This phenomenon appears clearly in Figure (\ref{fig:sub_iterations}) ("No sub resolution" label). When the delamination process propagates (Time steps 4 to 10), the number of LaTIn iterations to convergence becomes very large. A solution to this problem is developed in Section (\ref{sec:subiterations})
\end{itemize}



\section{Analysis of the iterative algorithm parameters}
\label{sec:ddr}
The search direction parameters $(k_E^+)_{(E \in \mathbf{E})}$ and $(k_E^-)_{(E \in \mathbf{E})}$ are introduced as positive definite symmetric operators. It has been empirically shown in previous studies that an optimal set of these operators exists. Though, these optimal operators are known to be difficult to interpret theoretically, especially when using complex interface behaviors. In addition, they can be expensive to compute even in simplified cases (perfect interface behavior). Our goal is here to find an efficient local approximation of the search direction operators for debonding analysis.

\subsection{Search direction $\mathbf{E}^+$}
\label{sec:kplus}

\begin{figure}[h]
       \centering
       \includegraphics[width=0.99 \linewidth]{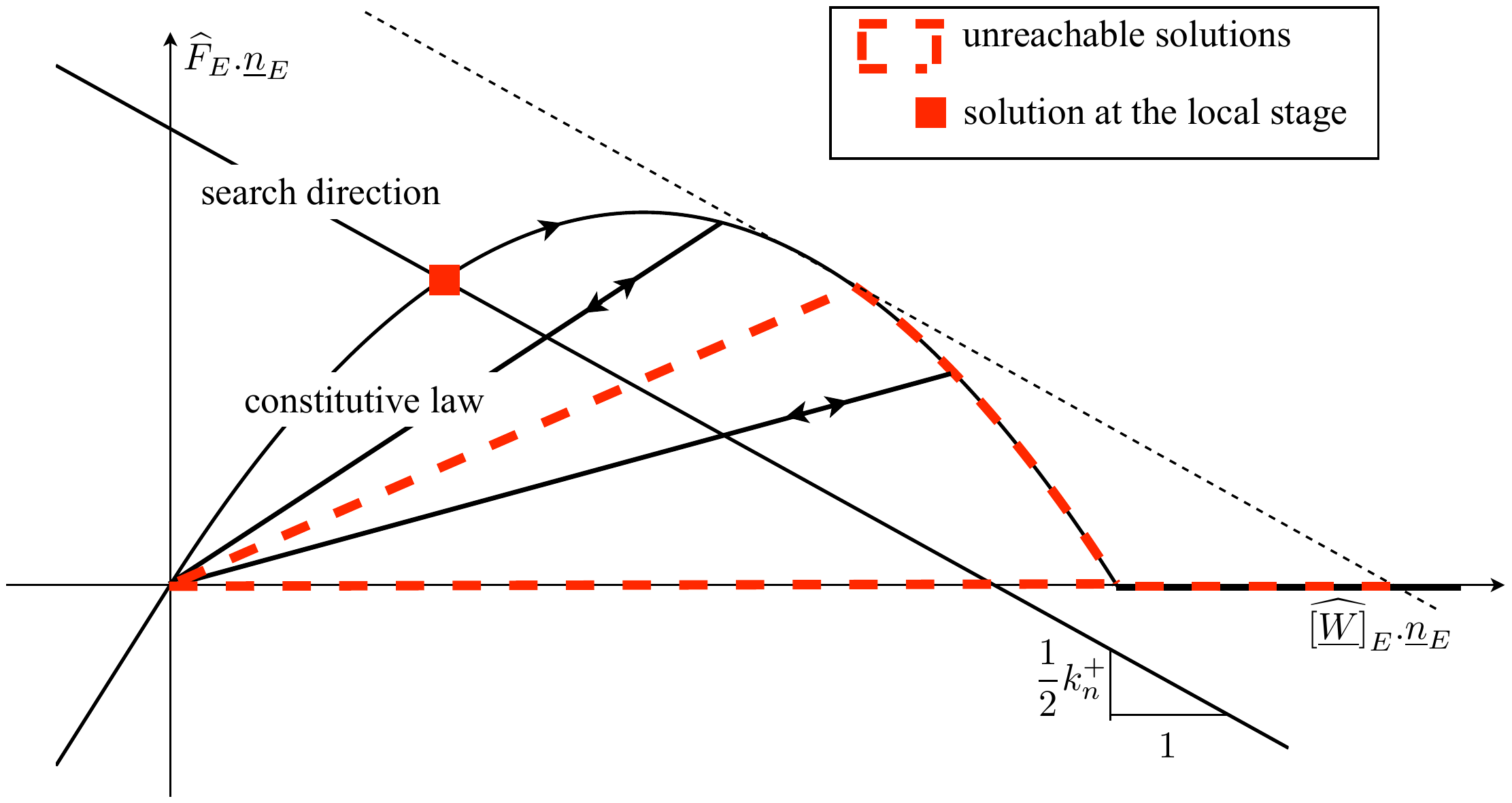}
       \caption{Using too small a value for Search direction parameter $k^+$}
       \label{fig:local_div}
\end{figure}

Using too small a value for  Parameters $(k_E^+)_{(E \in \mathbf{E})}$ can lead to the stagnation or divergence of the algorithm. Actually, in this case, the interface solution is seeked at the local stage in a truncated space $\boldsymbol{\Gamma}$, the solutions in the softening part of the local cohesive behaviour being unreachable. Figure (\ref{fig:local_div}) illustrates schematically this idea. The constitutive law of a cohesive interface and the search direction equation $\mathbf{E}^+$ are projected on the normal direction to the interface, at a given integration point. The solution seeked at the local stage is the intersection of the two curves obtained. This solution is computed by a Modified Newton-Raphson scheme. It clearly appears here that the part of the constitutive law within the dotted frame cannot be reached when using this iterative procedure. Thus, the global minimum of the problem may not be found through the LaTIn iterations.

\begin{figure}[h]
       \centering
       \includegraphics[width=0.99 \linewidth]{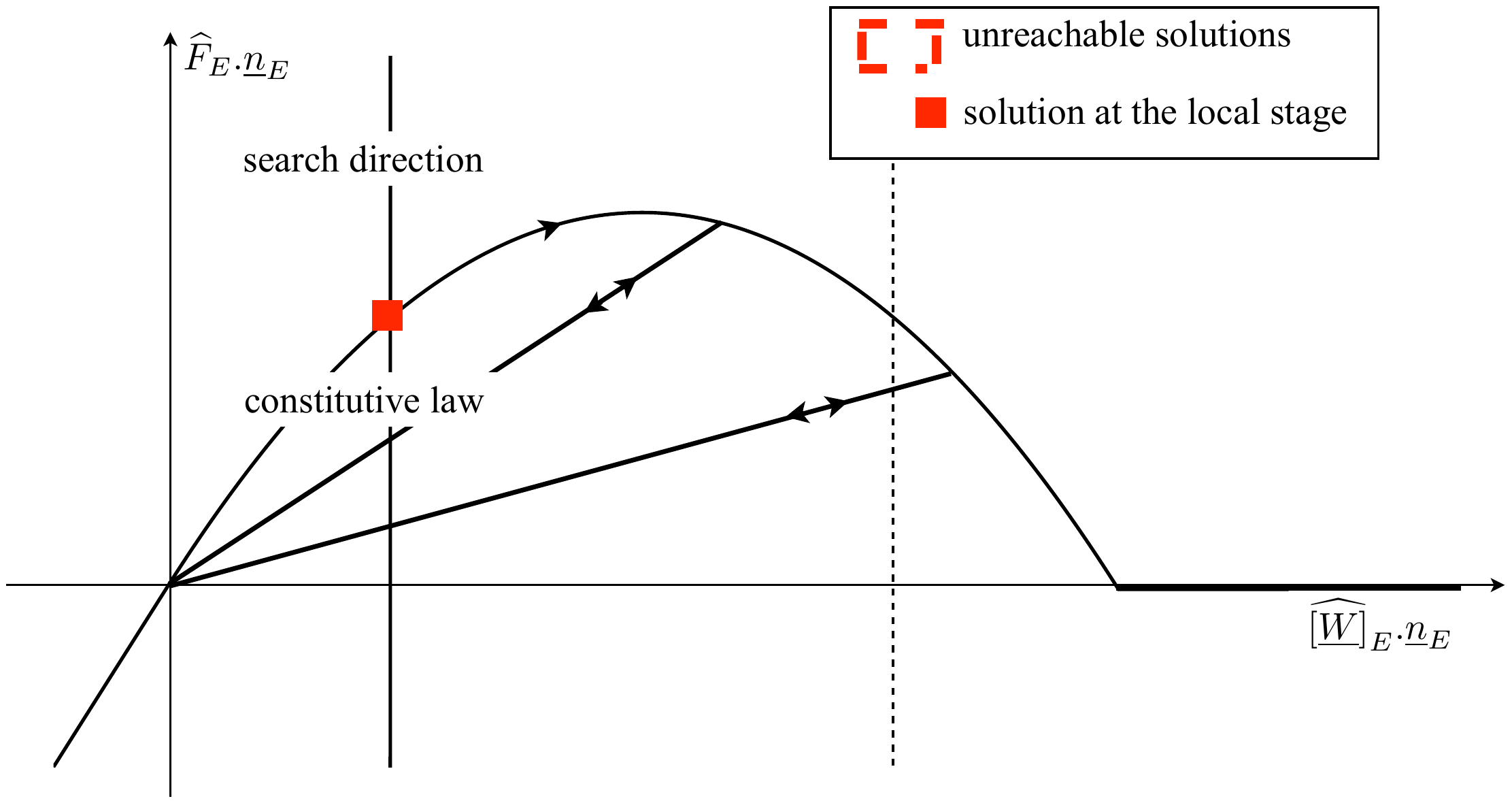}
       \caption{Using infinite value for Search direction parameter $k^+$}
       \label{fig:local_div2}
\end{figure}

We thus choose to set $(k_E^+)_{(E \in \mathbf{E})}$ to infinite values (see Figure \ref{fig:local_div2}) and focus on the choice of Search direction $\mathbf{E}^-$ only. Although slightly improved convergence rate could be obtained using classical conjugate search directions $\mathbf{E}^+$ and $\mathbf{E}^-$, the time costs of the local stage drop as the local problems can be solved directly.

\subsection{Search direction $\mathbf{E}^-$: interpretation}

The search direction $\mathbf{E}^-$ must be separated into a macro part ${\mathbf{E}^-}^\mathbf{M}$ and a micro part ${\mathbf{E}^-}^\mathbf{m}$ in order to be interpreted. Equation \eqref{eq:ddr_loc} can be rewritten:
\begin{equation}
\label{eq:ddr_macro_sep}
\begin{array}{ll}
\displaystyle \forall \ {\underline{W}_E^M}^\star & \displaystyle \in {\mathcal{W}_E^M}, \quad
\int_{\Gamma_{E}} \left( (\underline{F}_{E}-\underline{\widehat{F}}_{E})  \right)  . {\underline{W}_E^M}^\star \ d\Gamma
\\ 
& \displaystyle + \int_{\Gamma_{E}} \left( {k_E^-}^M \ (\underline{W}_{E}-\underline{\widehat{W}}_{E} - \underline{\widetilde{W}}^M \right) . {\underline{W}_E^M}^\star \ d\Gamma= 0
\end{array}
\end{equation}
\begin{equation}
\label{eq:ddr_micro_sep}
\begin{array}{ll}
\displaystyle \forall \ {\underline{W}_E^m}^\star & \displaystyle \in {\mathcal{W}_E^m}, \quad
\int_{\Gamma_{E}} \left( (\underline{F}_{E}-\underline{\widehat{F}}_{E}) \right) . {\underline{W}_E^m}^\star \ d\Gamma
\\ & 
\displaystyle  + \int_{\Gamma_{E}} \left( {k_E^-}^m \ (\underline{W}_{E}-\underline{\widehat{W}}_{E}) \right) . {\underline{W}_E^m}^\star \ d\Gamma= 0
\end{array}
\end{equation}
where ${\mathcal{W}_E^m}$ is the space orthogonal to ${\mathcal{W}_E^M}$ with respect to the $L^2(\Gamma_E)$ inner product.

\paragraph{Perfect interfaces}

Previous studies \cite{ladeveze00,ladeveze03b} have focused on the choice of the micro search direction parameter $({k_E^-}^m)_{E \in \mathbf{E}}$. Classically, when dealing with perfect interfaces, the optimal parameter ${k_E^-}^m$ on interface $\Gamma_E$ can be linked to the Schur complement of the structure occupying the domain $\Omega \setminus \Omega_E$. As the introduced microscopic interface quantities have a local influence, this Schur complement can be calculated in the vicinity of Interface $\Gamma_E$. In the case of isotropic and homogeneous materials, The scalar $E/L$ has been shown to be a good approximation of this local operator \cite{ladeveze03b}, $E$ being the Young's modulus of the adjacent substructure and $L$ a characteristic length of the interface. As explained in paragraphs, computing this microscopic optimal search direction parameter is not of primal interest in our case.

The $({k_E^-}^M)_{E \in \mathbf{E}}$ search direction parameter can be interpreted by considering an interface $\Gamma_{EE'}$ separating two adjacent substructures $E$ and $E'$. In order to simplify the explanations, we consider here a unique search direction parameters on $\Gamma_{EE'}$ for both substructures. Taking into account the macroscopic equilibrium of the interface forces along the interface $\Gamma_{EE'}$, on can derive from  \eqref{eq:ddr_macro_sep} :
\begin{equation}
\label{eq:soustr_ddr_macro}
\begin{array}{ll}
\displaystyle \forall  & \displaystyle {{\underline{W}}^M}^\star  \in {\mathcal{W}_{E}^M}, \quad
\\
& \displaystyle \int_{\Gamma_{EE'}} \left( \underline{F}_{E}-\frac{1}{2}  k_{EE'}^{-^M} (\underline{W}_{E'}-\underline{W}_{E} )  \right)   . {{\underline{W}}^M}^\star \ d\Gamma
\\
 & \quad  \displaystyle = \int_{\Gamma_{EE'}}  \left( \underline{\widehat{F}}_{E}-\frac{1}{2}  k_{EE'}^{-^M} (\underline{\widehat{W}}_{E'}-\underline{\widehat{W}}_{E} )
\right.)  . {{\underline{W}}^M}^\star \ d\Gamma
\end{array}
\end{equation}
This equation means that $k_{EE'}^{-^M}$ is a macroscopic stiffness of interface $\Gamma_{EE'}$. In the case of perfect interface, setting $k_{EE'}^{-^M}$ to the infinity equals to enforcing the macroscopic displacement continuity through the interface. As the the macroscopic equilibrium of the interface forces is ensured by equation \eqref{eq:ddr_macro}, this choice of the search direction parameter $k_{EE'}^{-^M}$ leads to the complete enforcement of the interface macroscopic behavior at the linear stage : 
\begin{equation} 
\label{eq:ddr_parfait_macro}
\left\{
\begin{array}{l}
\displaystyle \forall \ \displaystyle {\underline{W}_E^M}^\star \in {\mathcal{W}_E^M} \quad
\int_{\Gamma_{EE'}} \left( \underline{F}_{E}+\underline{F}_{E'} \right) . {\underline{W}_E^M}^\star \ d\Gamma = 0
\\  
\displaystyle \forall \ \displaystyle {\underline{W}_E^M}^\star \in {\mathcal{W}_E^M} \quad
\int_{\Gamma_{EE'}} \left( \underline{W}_{E}-\underline{W}_{E'} \right) . {\underline{W}_E^M}^\star \ d\Gamma = 0
\end{array} \right.
\end{equation}

Though, as the substructures are small compared to the size of the structure, the characteristic length introduced to approximate the microscopic optimal search direction parameter is very small. Hence, setting $k_{EE'}^{-^M}$ to $k_{EE'}^{-^m}$ is close to ensuring the macroscopic displacement continuity through perfect interfaces. Moreover, as it will be explained further on, the choice of a unique search direction paramater increases the simplicity and efficienty of the iterative strategy. \\

\begin{figure}[h]
       \centering
       \includegraphics[width=0.99 \linewidth]{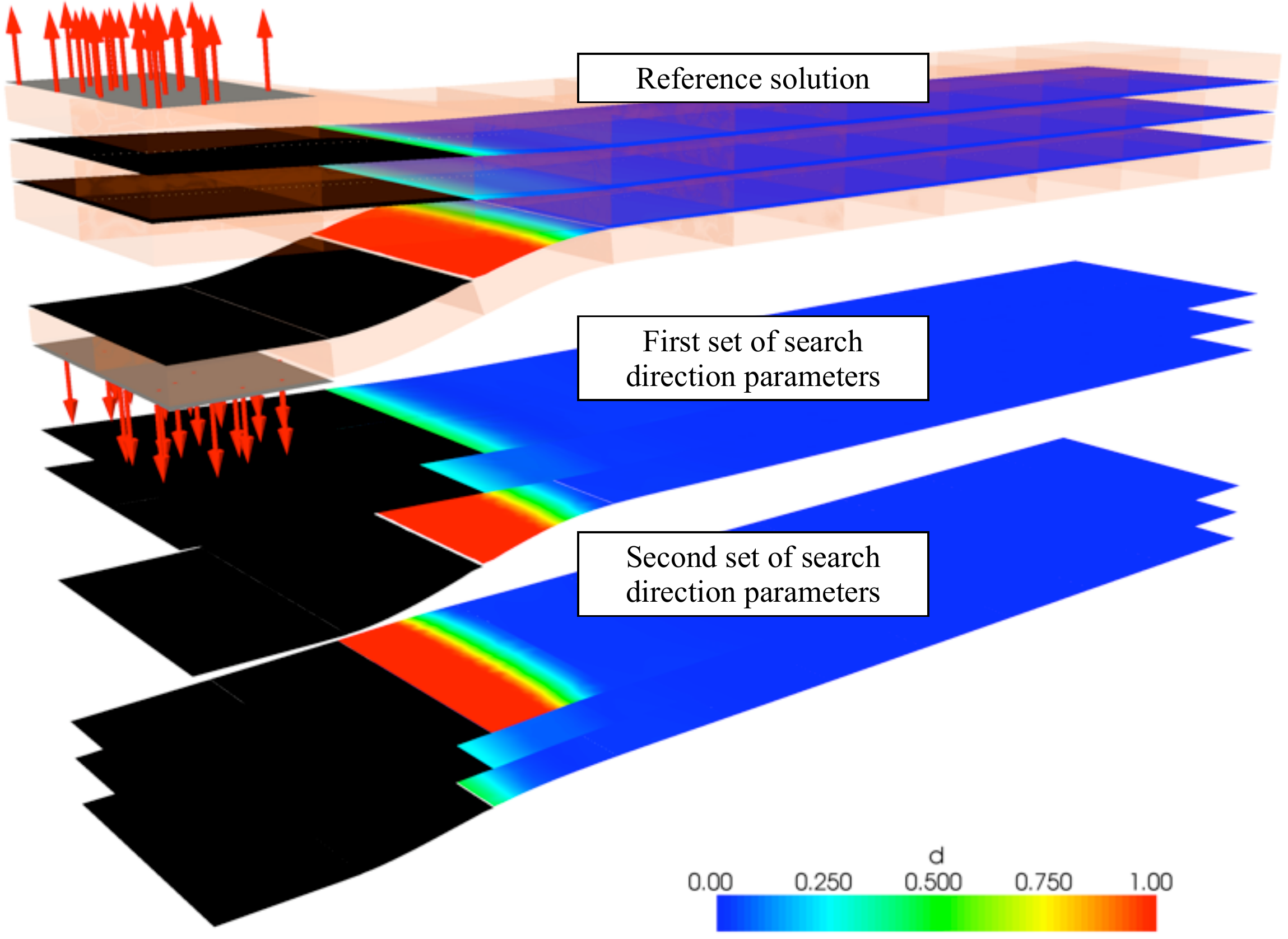}
       \caption{Converged solutions reached by the LaTIn algorithm using different sets of search direction parameters}
       \label{fig:non_physique}
\end{figure}

In the case of more complex interface behaviors, like softening, ignoring the influence of the macroscopic stiffness $({k_E^-}^M)_{(E \in \mathbf{E})}$ can lead to non physical solutions or even to the divergence of the iterative process. This idea is illlustrated by the results in Figure (\ref{fig:non_physique}). The test case presented is a four-plies composite beam, the black parts of the cohesive interfaces corresponding to initial delamination. The third interface is artificially weakened. The solution labelled "Reference solution" is obtained by a modified Newton algorithm, while the two following ones result from a LaTIn computation, two different set of search direction parameters being used.

Thus, an effort must be made to correctly set the interface conditions prescribed at the linear stage. In the case of simple interface behaviors (including the perfect interface behavior studied in the previous paragraph), a "physical" macroscopic interface condition can be prescribed beetween sub-structures at the linear stage. We will derive the general case from the analysis of these specific studies.

\paragraph{Homogeneous isotropic elastic interface}

Let us set the macroscopic search direction parameter $k_{EE'}^{-^M}$ of an elastic interface $\Gamma_{EE'}$ to $2 \, k^0 (1-d)$,  where $k^0$ is the local scalar stiffness of the interface. Equation \eqref{eq:soustr_ddr_macro} now reads :
\begin{equation}
\left\{ \begin{array}{ll}
\displaystyle \forall  & \displaystyle {\underline{W}_E^M}^\star \in {\mathcal{W}_E^M} \quad
\int_{\Gamma_{EE'}} \left( \underline{F}_{E}+\underline{F}_{E'} \right) . {\underline{W}_E^M}^\star \ d\Gamma = 0 \\

\displaystyle \forall & \displaystyle {\underline{W}_{EE'}^M}^\star \in {\mathcal{W}_E^M},
\\ 
&\displaystyle \int_{\Gamma_{EE'}} \left( \underline{F}_{E} - k^0(1-d) \ (\underline{W}_{E'}-\underline{W}_{E}) \right) . {\underline{W}_E^M}^\star \ d\Gamma= 0
\end{array} \right.
\end{equation}
Hence, setting $\frac{1}{2} k_{EE'}^{-^M}$ to the interface stiffness equals to prescribing the macroscopic interface behavior as an interface condition at the linear stage.

\paragraph{Delaminated interfaces under traction}
In the case of a completely damaged cohesive interface loaded with traction (so that the gap of interface displacements is positive), the converged interface force fields are null. Thus, both micro and macro search direction parameters should optimaly be set to zero. Indeed, $\underline{\widehat{F}}_{E} = \underline{\widehat{F}}_{E'} = 0$ on interface $\Gamma_{EE'}$ as these quantities result from the local stage and verify the interface behavior. Consequently, setting $k_{EE'}^{-^M} = k_{EE'}^{-^m} = 0$ leads to the enforcement of the relation $\underline{F}_{E} = \underline{F}_{E'} = 0$ (equations \eqref{eq:ddr_macro_sep} and \eqref{eq:ddr_micro_sep}).
%


\begin{figure*}
       \centering
       \includegraphics[width=0.75 \linewidth]{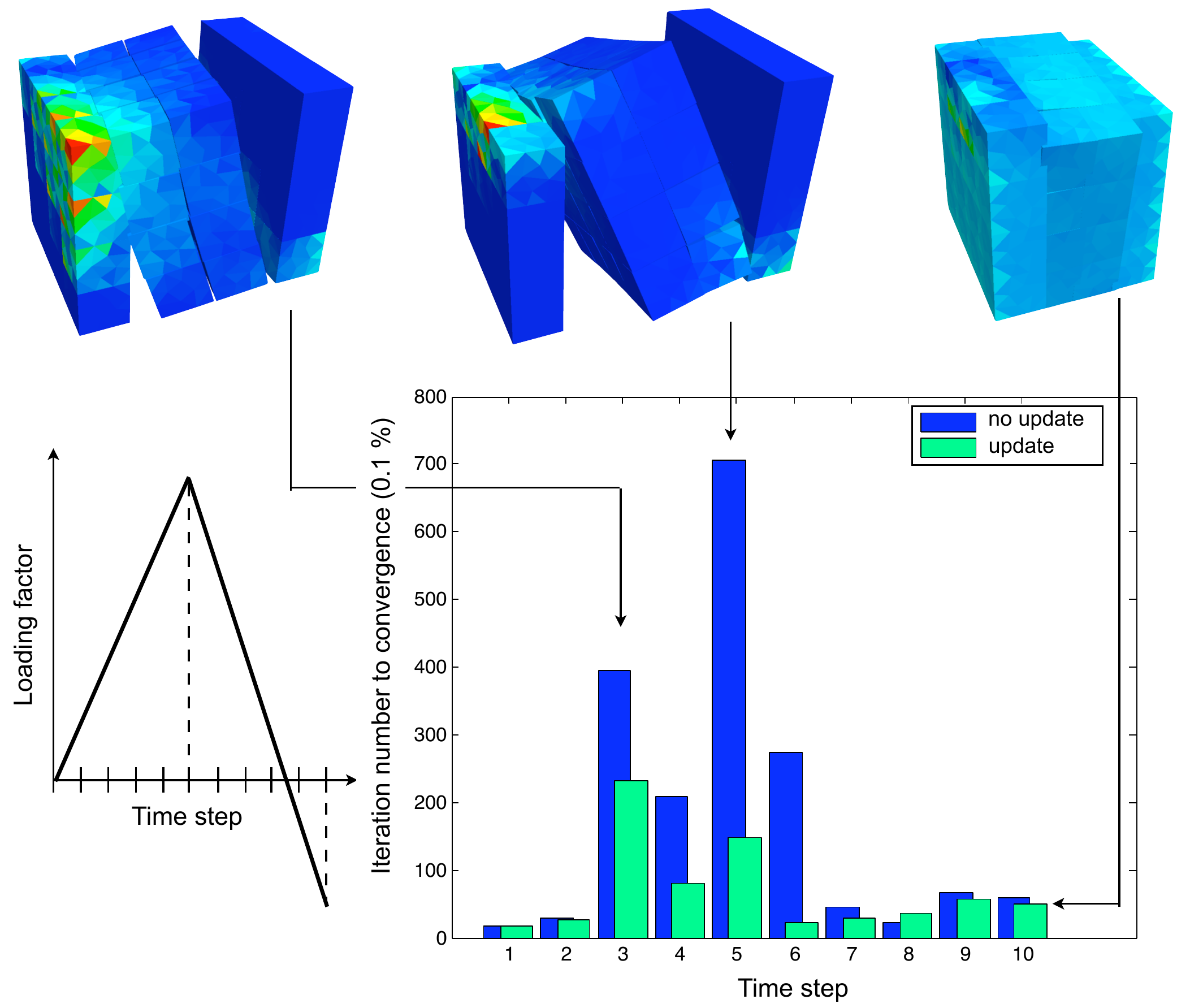}
       \caption{Use of small search direction parameters for delaminated interfaces}
       \label{fig:conv_pen}
\end{figure*}

\paragraph{General case} We can conclude from these three simplified cases that, as the macroscopic equilibrium of the interface forces is enforced at the linear stage through the choice of Admissibility space $\mathbf{A_d}$, one can ensure the verification of the complete linearized interface behavior by the macroscopic interface fields at this stage by setting the $\mathbf{E}^-$ interface parameters to specific values. Thus, together with the choice of a rigid $\mathbf{E}^+$ search direction, this choice leads to a hybrid iterative strategy :
\begin{itemize}
\item a modified Newton-Raphson scheme on the macro-part of the solution
\item a Latin-type algorithm on the micro part, the search directions $\mathbf{E}^+$ and $\mathbf{E}^-$ being non-conjugated (which is not classical)
\end{itemize}

Though, when dealing with cohesive interfaces with non-constant stiffness, with interfaces partially delaminated, or with delaminated interfaces loaded with both local traction and local compression, the interface behavior that should be verified by the macroscopic fields is not explicit. Yet a too coarse approximation of this behavior might drive the algorithm towards a local minimum of potential energy instead of a "more physical" global minimum. 

Moreover, the mechanical interpretation of the search direction parameters $(k_E^-)_{(E \in \mathbf{E})}$ requires to introduce a micro search direction ${\mathbf{E}^-}^\mathbf{m}$ and a macro search direction ${\mathbf{E}^-}^\mathbf{M}$. But when using this search direction separation, we found out that the CPU time increased, because of the large amount of projections of the interface fields in the macrospace required through the iterative process.

A practical way to choose a common micro and macro search direction parameter that ensures a good interface condition is the subject of next subsection.

\subsection{Search direction $\mathbf{E}^-$: practical choice}

The $(k_E^-)_{(E \in \mathbf{E})}$ search direction parameters are set with respect to the interface behavior, as explained bellow:
\begin{itemize}
\item \textit{perfect interfaces:} $(k_E^-)_{(E \in \mathbf{E})}$ are set to the optimal micro values described previsouly. As the characteristic length $L$ of the interface is very small, these values are high enough to ensure the continuity of the macrodisplacement through the perfect interfaces.
\item \textit{undamaged or partially damaged cohesive interfaces:} $(k_E^-)_{(E \in \mathbf{E})}$ are set to the optimal macro values for initially undamaged interfaces. In the case of an interface $\Gamma_{EE'}$, we thus choose:
\begin{equation}
k^-_{EE'} = 
\left(
\begin{array}{ccc}
  2\, k^0_n & 0 & 0 \\
  0 & 2\, k^0_t & 0 \\
  0 & 0 & 2\, k^0_t
\end{array}
\right)_{(\V{n}_E,\V{t}_{1},\V{t}_{2})}
\end{equation}
even though it cannot be shown theoritically, this strategy has always led to the convergence of the iterative algorithm. 
Moreover, as the interface conditions between substructures are connected to mechanical properties, no branching to non-physical solutions has been observed in our cases.
\item \textit{delaminated interfaces:} $(k_E^-)_{(E \in \mathbf{E})}$ are set to the same values used for undamaged interfaces, unless the contact status of the interface is known (i.e.: unless all integration points are in compression, or unless all integration points are in traction). In the last cases, the search direction parameters are updated as follows :

\begin{itemize}
\item[$\bullet$] delaminated interface under compression
\begin{equation}
k^-_{EE'} = 
\left(
\begin{array}{ccc}
  2\, k^0_n & 0 & 0 \\
  0 & 0 & 0 \\
  0 & 0 & 0
\end{array}
\right)_{(\V{n}_E,\V{t}_{1},\V{t}_{2})}
\end{equation}

\item[$\bullet$] delaminated interface under traction
\begin{equation}
k^-_{EE'} = 0 \times \operatorname{I_d}
\end{equation}
\end{itemize}
Obviously, updating $\mathbf{E}^+$ with respect to the contact status of the cohesive interface requires a re-assembling step of the macroscopic global operator. Potentially, this method can be expensive, unless the macroscopic problem is solved using a parallel strategy (see Section \ref{sec:third_scale}).

\begin{figure*}
       \centering
       \includegraphics[width=0.85 \linewidth]{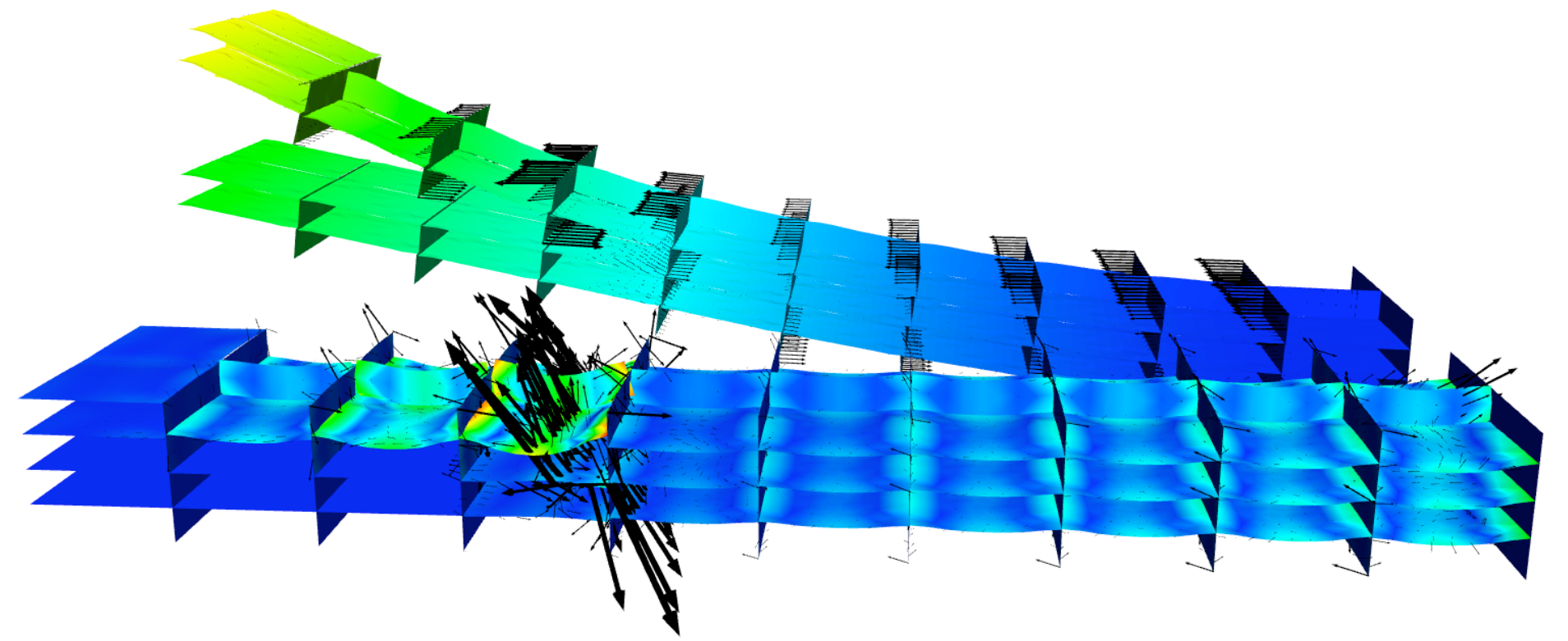}
       \caption{Micro interface quantities $\times 10$ (front)
       and macro interface quantities (back)}
       \label{fig:DCB_mM}
\end{figure*}

The results of this procedure for the cubic test case represented Figure (\ref{fig:DCB}) are shown in Figure (\ref{fig:conv_pen}). All the interfaces between adjacent sub-structures are granted a cohesive behavior. The prescribed loading leads to initiation of the delamination (time steps 1 to 2), opening of the cracks (time steps 3 to 7) and closing of the cracks (time steps 8 to 10). In the first case, the search direction parameters $(k_E^-)_{(E \in \mathbf{E})}$ are constant throughout the analysis. In the second case, they are updated as explained previously, the contact status of the delaminated interfaces being checked every ten LaTIn iterations. Clearly, the number of iterations to convergence is significantly reduced when delamination occurs, the delaminated interface being under traction.
\item \textit{prescribed forces (respectively displacement) interfaces:} $(k_E^-)_{(E \in \mathbf{E})}$ are set to a very low (respectively high) value so as to enforce the boundary condition through penalization to the adjacent substructure.
\end{itemize}

In the end, it comes out that the retained parameters are quite independant on the shape of the cohesive law. Thus, good performance results can be expected from other damaging interface models, without significant adaptation.

\section{Subresolutions in the crack's tip vicinity}
\label{sec:subiterations}
\begin{figure*}
       \centering
       \includegraphics[width=0.8 \linewidth]{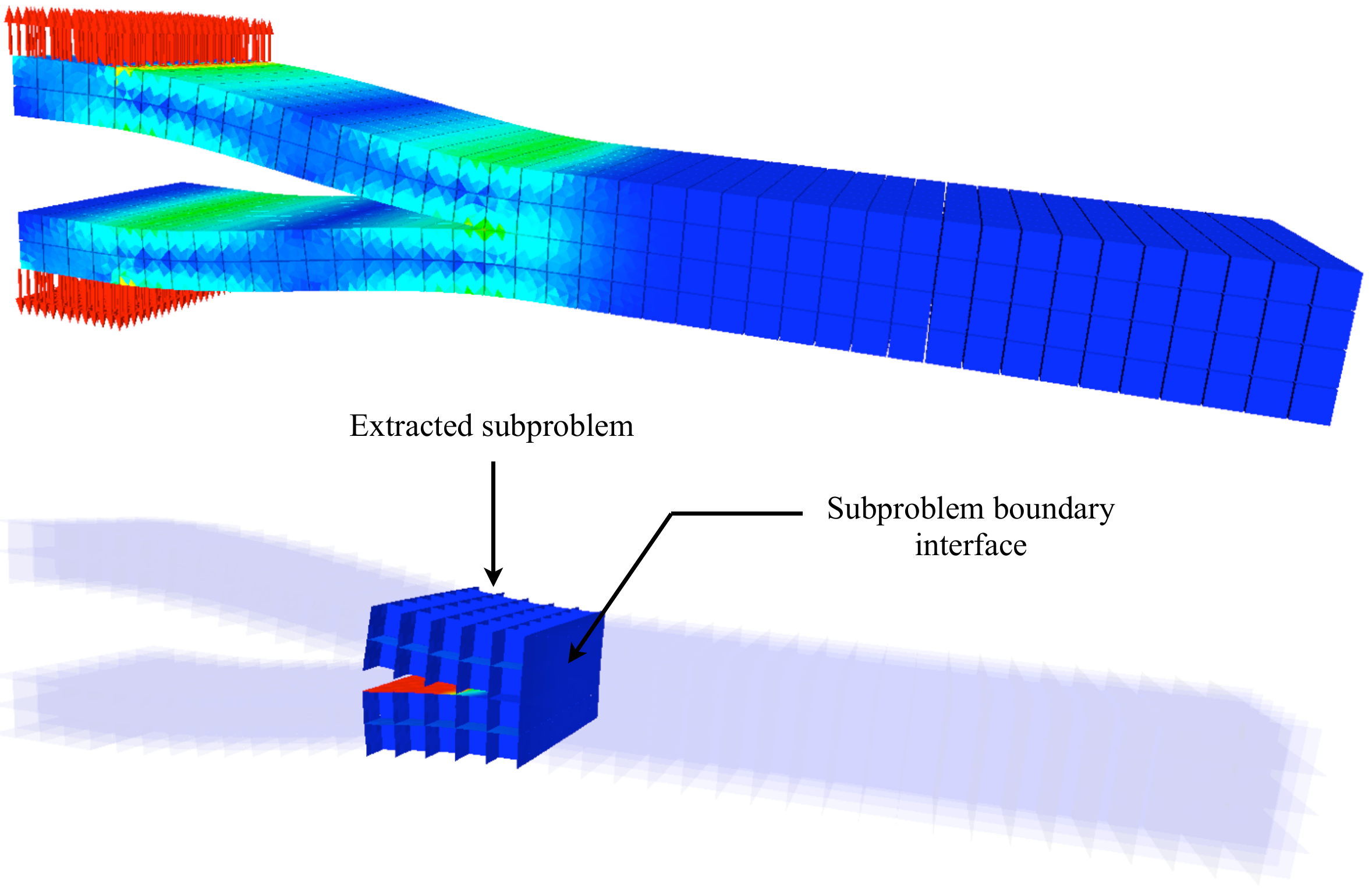}
       \caption{Extraction of a subproblem}
       \label{fig:adap_endom}
\end{figure*}

The drop in the convergence rate occurring when the cracks propagate can be explained by two main phenomena:
\begin{itemize}
\item the singularity near the tip of the crack is very poorly represented by linear macro quantities (see Figure (\ref{fig:DCB_mM})). Therefore the complementary ``micro'' parts of these phenomena, which are calculated iteratively through the resolution of local problems, have an influence on a significant part of the structure. In order to maintain the scalability of the method, 
a method to filter this global influence from the ``micro''  quantities in the process zone and then transmit it to the whole structure must be designed.
\item the prediction of the location of the crack's tip at a given time step requires the quasi-convergence of a large number of consecutive equilibrium states. Consequently, the propagation of the crack is very slow as the iterations proceed.
\end{itemize}
A first solution would be to enlarge the macro space so that the totality of the numerical influence of the crack would be systematically transmitted over the whole structure. Because such a strategy would imply to update (reassembly and refactorization) the macro problem at each evolution of the crack, it is not computationally realistic. However, the very-large-variation-length part of the solution is determined correctly in most of the structure since the very first iterations. Therefore as exposed in this section, one can choose to solve ``exactly'' the highly nonlinear problem in the crack's front vincinity at each global LaTIn iteration.

\subsection{Principle of the sub-resolution strategy}

The technique consists in extracting a part $\Omega_{sub}$ of Domain $\Omega$. The converged solution of the extracted nonlinear subproblem on $\Omega_{sub}$ is sought using the two-scale domain decomposition strategy described in Section \ref{sec:reference_problem} (see Figure (\ref{fig:adap_endom})) along with Algorithm (\ref{alg:sub_iter})). This idea is similar to the concept of nonlinear relocalization developed in \cite{cresta07,pebrel08}, in which the authors used a domain decomposition method and performed a nonlinear analysis in each substructure after a resolution of the condensed global linearized problem. In our case, the nonlinear relocalization is carried out on a set of substructures because the nonlinearities are localized at the interfaces of the domain decomposition scheme.

Let us concentrate on two main difficulties:
\begin{itemize}
\item the choice of the boundary conditions for the subproblem
\item the choice of the size and position of the extracted subdomain
\end{itemize}

\subsection{The boundary conditions of the subproblem}

The subresolution is carried out at each step of the global iterative resolution, which results in unbalanced substructure and interface fields. Numerical tests showed that the prescribed conditions applied on the boundary $\partial \Omega_{sub}$ of domain $\Omega_{sub}$ must be in Space $A_d$ (\textit{i.e.} they must result from a linear stage of the resolution). Indeed these fields are in global static equilibrium over the whole structure, whereas the solutions in Space $\Gamma$ are in equilibrium only locally and do not self-equilibriate $\Omega_{sub}$.

Thus, subresolutions can be interpreted as an enhancement of the linear stage. In order the process zone to keep matching the remaining of the structure, Robin boundary conditions are prescribed on $\partial\Omega_{sub}$ using search direction $\mathbf{E}^-$ as interface stiffness parameter.

\subsection{Adaptivity of the subproblem}

In order to extract the subproblem automatically, we choose to select a set of substructures and interfaces in a box surrounding the interface with the highest damage rate at the end of the global step (see Figure (\ref{fig:adap_endom})).

\begin{figure}
        \centering
        \includegraphics[width=0.99 \linewidth]{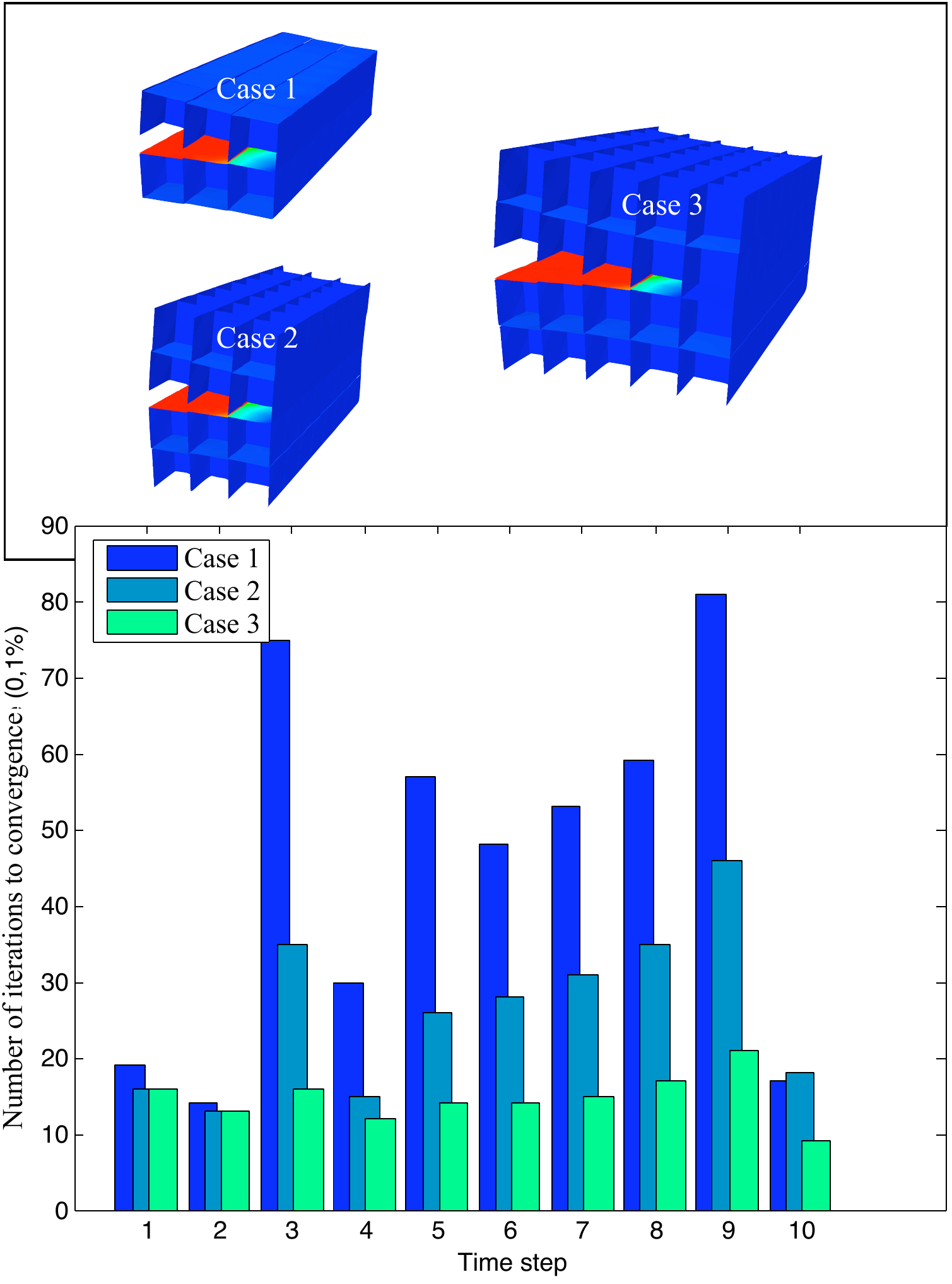}
    \caption{Influence of the size of the subproblem}
    \label{comp_size}
\end{figure}

The influence of the size of the extracted subdomain is shown in Figure (\ref{comp_size}).

\begin{algorithm}[ht]\caption{The subresolution strategy algorithm}\label{alg:sub_iter}
\begin{algorithmic}[1]
\STATE Substructures' operators construction
\STATE Computation of the macroscopic homogenized behavior $\mathbb{L}_E^M$ on each substructure 
\STATE Global assembly of the macroscopic operator $\mathbf{L^M}$
\STATE Initialization $\mathbf{s}_0 \in \boldsymbol{\Gamma}$
\FOR{$n=0,\ldots,N$}%
  \STATE Linear stage: computation of $\mathbf{s}_n \in \mathbf{A_d}$ \\
  \STATE Local stage: computation of $\mathbf{\widehat{s}}_{n+\frac{1}{2}} \in \boldsymbol{\Gamma}$ \\
  \STATE Subproblem extraction\\
  $\quad \square$ Location of the substructures requiring subresolution \\
  $\quad \square$ Application of mixed boundary conditions \\
  $\quad \square$ Assembly of the macro subproblem
  \FOR{$j=0,\ldots,m$} %
    \STATE Subproblem linear stage
    \IF{$j \leq m-1$}
    \STATE Subproblem local stage
    \ENDIF
    \STATE Local error indicator
  \ENDFOR
  \STATE Local stage on the boundary interfaces of the subproblem
  \STATE Calculation of an error indicator
\ENDFOR
\end{algorithmic} \end{algorithm}

\subsection{Results}

The resolution of a subproblem around the crack's tip leads to a convergence rate of the global resolution which is independent of the time step of the analysis (\textit{i.e.} independent of the area of the interface which becomes delaminated in one time step), which means that the numerical scalability is restored.

As a result, the local inversion time for the problem shown in Figure (\ref{fig:adap_endom}) (with the smaller subiteration domain) was cut in half. This estimate does not take into account the fact that the macroproblem is much smaller in the case of the subiterations; furthermore the gain increases as the ratio between the size of the process zone and the size of the structure decreases.

Thus, this method can lead to a reduction in the number of calculations. However, this reduction would be ineffective unless the subproblem is re-parallelized. Indeed, using the initial allocation among the parallel processors would adress the extracted subproblem to only a very small number of processors. Another solution, easier to implement but potentially less efficient, would be to perform independent subiterations systematically on all the processors.

\section{Third-scale resolution}
\label{sec:third_scale}

\begin{figure*}
\centering
\subfigure[Stresses in substructures]{\includegraphics[width=0.72 \linewidth]{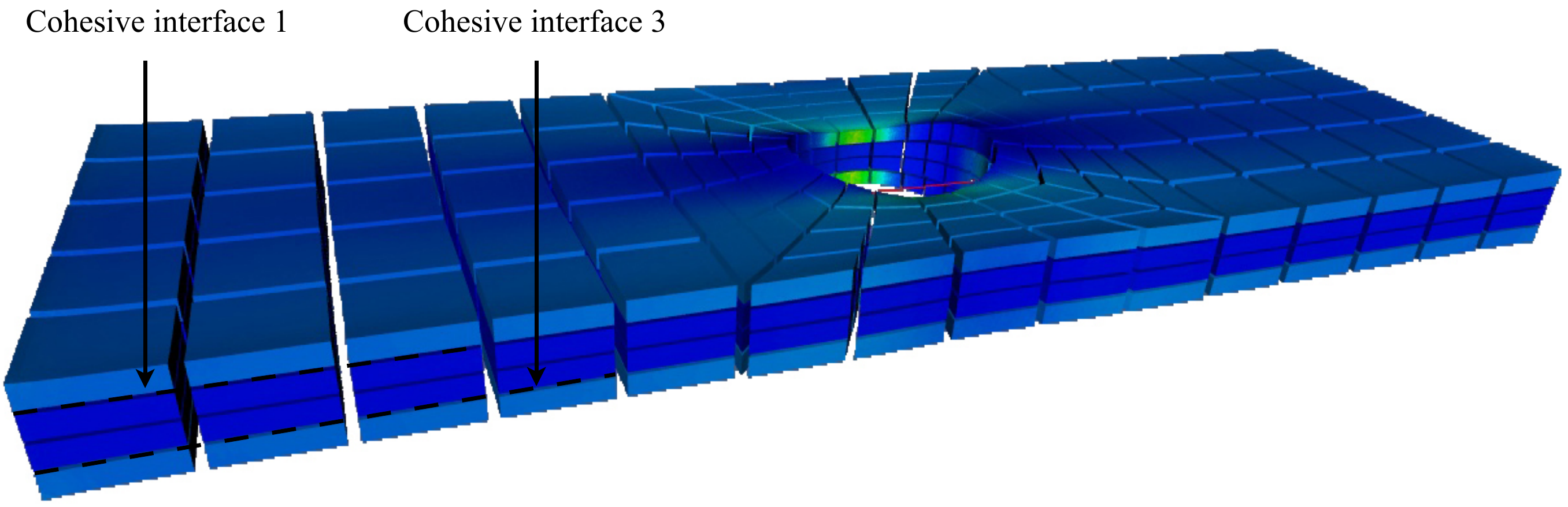}}
\subfigure[Damage in Cohesive interface 1]{\includegraphics[width=0.7 \linewidth]{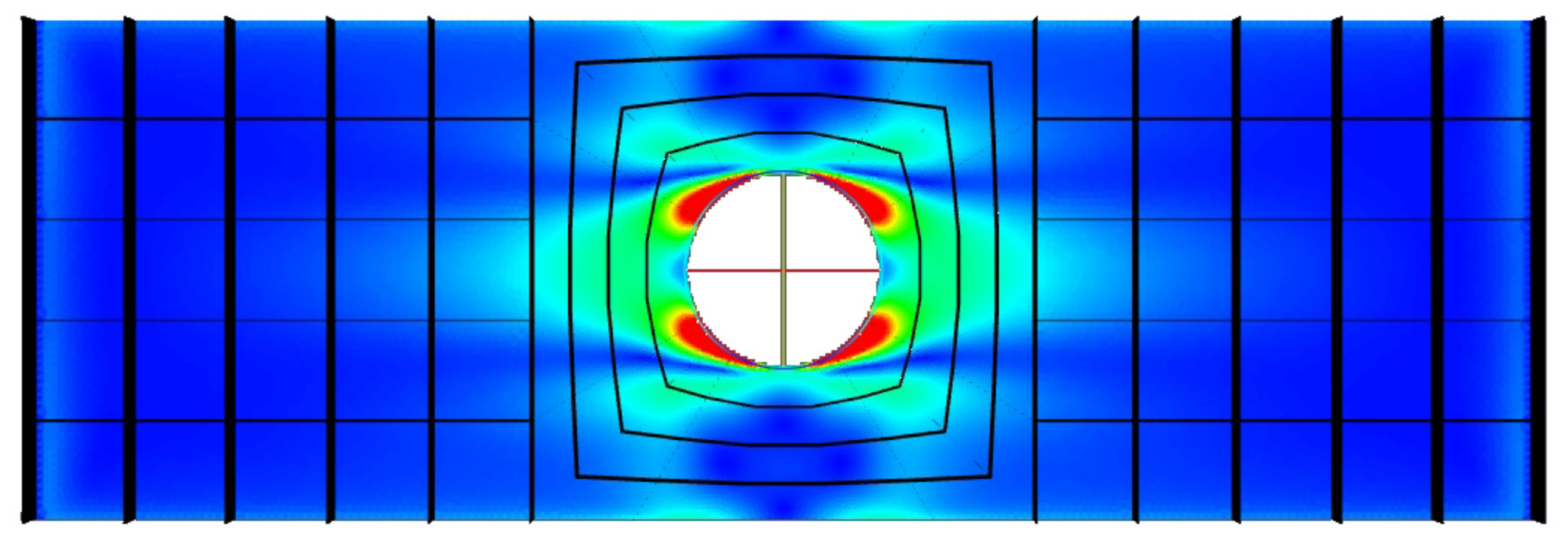}}
\caption{The four-ply holed plate problem (3.4 Mdof)}
\label{fig:holed}
\end{figure*}

The substructuring described in Section \ref{sec:reference_problem} results in a very large macro problem and in an unnecessarily refined macroscopic solution. In order to solve large problems such as the one represented Figure (\ref{fig:holed}), we need to focus on:
\begin{itemize}
\item the parallel resolution of the macroproblem,
\item the selection and transmission of the large-wavelength part of the macro
solution.
\end{itemize}

These two elements can be introduced into the method through the use
of any Schur-complement-based domain decomposition method
\cite{gosselet06}. We chose to implement the BDD method \cite{mandel93} to solve the macroproblem.

\subsection{The balancing domain decomposition method for the
macroproblem}

\subsubsection{Partitioning of the macroproblem}

\begin{figure*}
       \centering
       \includegraphics[width=0.72 \linewidth]{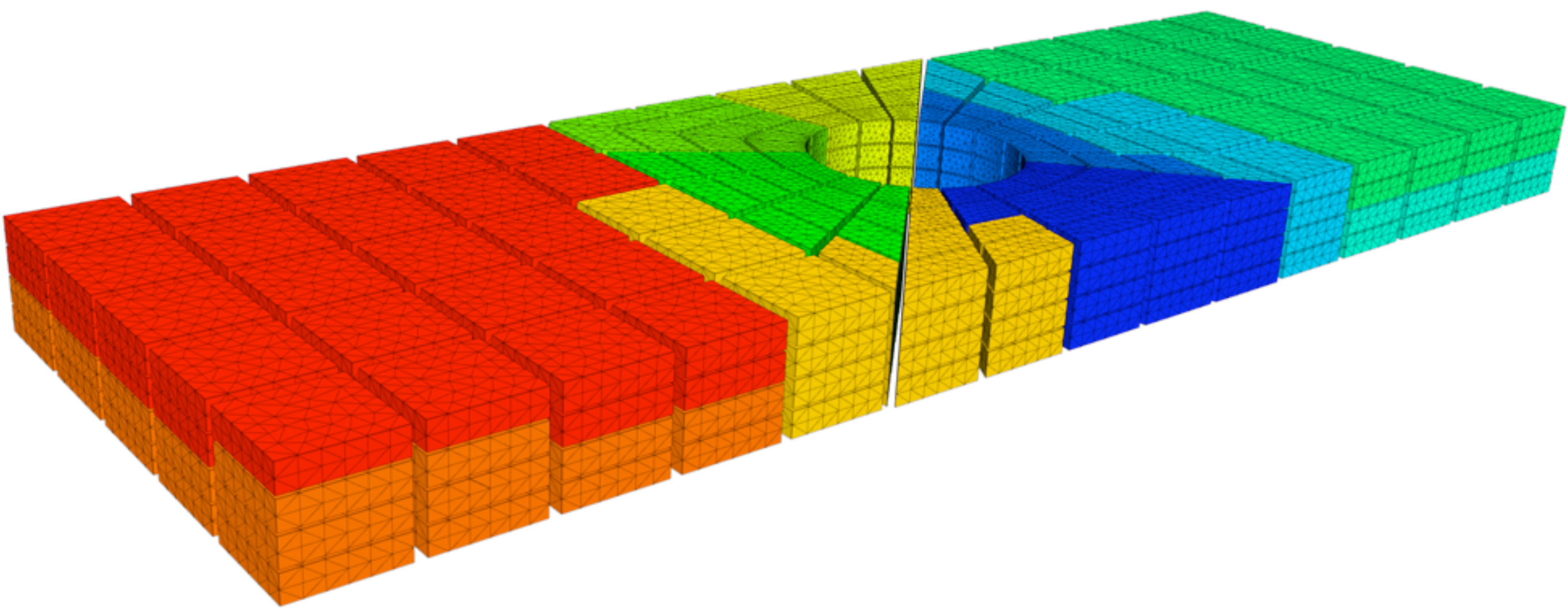}
       \caption{3-level substructuring:  substructures assignment to processors}
       \label{fig:third_scale}
\end{figure*}

The substructures of the initial partitioned problem are grouped
into super-substructures $\bar{E}$ separated by super-interfaces
$\Gamma_{\bar{E}\bar{E'}}$ (Figure (\ref{fig:third_scale})). Practically, each super-sub-structure is made of the whole set of sub-structures assigned to a given processor of the parallel computing architecture. The
algebraic problem to be solved within each of these super-substructures reads:

\begin{equation}
\label{eq:decomp}
\left\{ \begin{array}{c}
\left( \begin{array}{cc}
\displaystyle \mathbf{\mathbf{L}^M}_{ii}^{(\bar{E})} & \displaystyle \mathbf{\mathbf{L}^M}_{ib}^{(\bar{E})} \\
\displaystyle \mathbf{\mathbf{L}^M}_{bi}^{(\bar{E})} & \displaystyle \mathbf{\mathbf{L}^M}_{bb}^{(\bar{E})}
\end{array} \right)
\left( \begin{array}{c}
\displaystyle \widetilde{W}_i^{(\bar{E})} \\
\displaystyle \widetilde{W}_b^{(\bar{E})}
\end{array} \right)
=
\left( \begin{array}{c}
\displaystyle F_i^{(\bar{E})} \\
\displaystyle F_b^{(\bar{E})}+ \widetilde{\lambda}_b^{(\bar{E})}
\end{array} \right)
\\
\displaystyle \sum_{\bar{E}}  \mathbf{B}^{(\bar{E})}  \widetilde{W}_b^{(\bar{E})} = 0
\\
\displaystyle \sum_{\bar{E}}  \mathbf{A}^{(\bar{E})}  \widetilde{\lambda}_b^{(\bar{E})} = 0
\end{array} \right.
\end{equation}
where the $^M$ superscript has been omitted, the subscripts $b$ and
$i$ refer to the super-interface quantities and to the internal
quantities of the super-substructures respectively.
$\mathbf{B}^{(\bar{E})}$ and $\mathbf{A}^{(\bar{E})}$ are signed
Boolean localization operators. The second equation of System
\eqref{eq:decomp} expresses the continuity of the kinematic
unknowns, while the third equation expresses the equilibrium of the
nodal reactions at the interfaces between super-substructures.

The local equilibria are condensed at the super-interfaces through
the introduction of Schur complements $\mathbf{S}^{(\bar{E})}$ and
 condensed forces $\underline{F}_c^{(\bar{E})}$:
\begin{equation}
\label{eq:cond_sst}
\mathbf{S}^{(\bar{E})} \ \widetilde{W}_b^{(\bar{E})} = \widetilde{\lambda}_b^{(\bar{E})}+ F_c^{(\bar{E})}
\end{equation}
\begin{equation}
\nonumber
\textrm{where} \quad
\left\{ \begin{array}{c}
\displaystyle \mathbf{S}^{(\bar{E})} = \mathbf{\mathbf{L}^M}_{bb}^{(\bar{E})} - \mathbf{\mathbf{L}^M}_{bi}^{(\bar{E})} \  \mathbf{\mathbf{L}^M}_{ii}^{{(\bar{E})}^{-1}} \ \mathbf{\mathbf{L}^M}_{ib}^{(\bar{E})} \\
\displaystyle F_c^{(\bar{E})} = F_b^{(\bar{E})} - \mathbf{\mathbf{L}^M}_{bi}^{(\bar{E})} \  \mathbf{\mathbf{L}^M}_{ii}^{{(\bar{E})}^{-1}} F_i^{(\bar{E})}
\end{array} \right.
\end{equation}

The continuity of displacement is achieved automatically through the
introduction of a unique super-interface macrodisplacement
$\underline{\widetilde{W}}_b$. Then, the continuity equation of the
interface reaction forces yields:
\begin{equation}
\mathbf{S} \  \widetilde{W}_b = F_c
\end{equation}
\begin{equation}
\nonumber
\textrm{where} \quad
\left\{ \begin{array}{c}
\displaystyle \mathbf{S} = \sum_{\bar{E}}  \mathbf{A}^{(\bar{E})} \mathbf{S}^{(\bar{E})} {\mathbf{A}^{(\bar{E})}}^T \\
\displaystyle  F_c = \sum_{\bar{E}}  \mathbf{A}^{(\bar{E})} \underline{F}_c^{(\bar{E})}
\end{array} \right.
\end{equation}

\subsubsection{Resolution of the super-interface problem}

The condensed macroproblem is solved iteratively through a conjugate
gradient algorithm. Classically, this resolution requires only
matrix-vector products and dot products, which are compatible with
parallel architectures.

\begin{algorithm}[ht]\caption{Projected preconditioned conjugate gradient}\label{alg:cg:1}
\begin{algorithmic}[1]
\STATE Initialize $\widetilde{W}_{b_0} = (\proj
\widetilde{W}_{b_{00}}) + \mathbf{C} (\mathbf{C}^T \mathbf{S}
\mathbf{C})^{-1} \mathbf{C}^T F_c$ \STATE Calculate
$r_0=F_c-\mathbf{S} \widetilde{W}_{b_0}$
\STATE Calculate $z_0 = \proj \mathbf{\widetilde{S}^{-1}} r_0$ and set $w_0=z_0$%
\FOR{$j=0,\ldots,m$} %
  \STATE $\alpha_j=(r_j,z_j)/(\mathbf{S} w_j,w_j)$%
  \STATE $\widetilde{W}_{b_{j+1}} =\widetilde{W}_{b_j}+\alpha_j w_j$%
  \STATE $r_{j+1}=r_j-\alpha_j \mathbf{S} w_j$%
  \STATE $z_{j+1} = \proj \mathbf{\widetilde{S}}^{-1} r_{j+1}$
  \STATE $\beta^j_j=-(\mathbf{S} w_{j},z_{j+1})/(w_j,\mathbf{S} w_j)$
  \STATE $w_{j+1}=z_{j+1}+\sum_{k=0}^j\beta^k_j w_k$%
\ENDFOR
\end{algorithmic} \end{algorithm}

The recommended preconditioner for a parallel use of this algorithm
is what is called the Neumann preconditioner:
\begin{equation}
\mathbf{\widetilde{S}}^{-1} =
\sum_{\bar{E}} \mathbf{A}^{(\bar{E})} {\mathbf{S}^{(\bar{E})}}^{+} {\mathbf{A}^{(\bar{E})}}^T
\end{equation}
${\mathbf{S}^{(\bar{E})}}^{+}$ being a generalized inverse of the
Schur complement of Super-substructure $E$.

\noindent The use of this preconditioner means that the inverse of
the global super-macro operator is approximated by the assembly of
the inverses of the local Schur complements. Let us note that the
description chosen for the interface macrofields precludes the
existence of degrees of freedom belonging to more than two
substructures; consequently, no scaling is required in the
preconditioner (at least when the interfaces are not too
heterogeneous).

Since the product $\mathbf{\widetilde{S}}^{-1} r_{j+1}$ consists in solving
Neumann problems for each super-substructure $\bar{E}$ under the
loading $r_{j+1}$, one must ensure that $r_{j+1}$ is
self-balanced in the sense of $\bar{E}$. Therefore, we introduce
a projector $\mathbf{P}$ which projects the residual onto the space
orthogonal to the kernel of the super-substructure at each iteration
of the conjugate gradient.

Thus, the solution is sought in the form:
\begin{equation}
\widetilde{W}_b = \widetilde{W}_{b_0} + \mathbf{P} \ \widetilde{\widetilde{W}}_b
\end{equation}
\begin{equation}
\nonumber
\begin{array}{l}
\textrm{where} \\
\left\{ \begin{array}{rcl}
\displaystyle \mathbf{C}^T (b - \mathbf{S} \widetilde{W}_{b_0}) = 0 & \displaystyle \Longrightarrow & \displaystyle \widetilde{W}_{b_0} = \mathbf{C} (\mathbf{C}^T \mathbf{S} \mathbf{C})^{-1} \ \mathbf{C}^T F_c \\
\displaystyle \mathbf{C}^T \mathbf{S} \mathbf{P} = 0 & \displaystyle \Longrightarrow & \displaystyle \mathbf{P} = \mathbf{I } - \mathbf{C}(\mathbf{C}^T \mathbf{S} \mathbf{C})^{-1} \mathbf{C}^T \mathbf{S}
\end{array} \right.
\end{array}
\end{equation}

\begin{figure}[t]
       \centering
       \includegraphics[width=0.99 \linewidth]{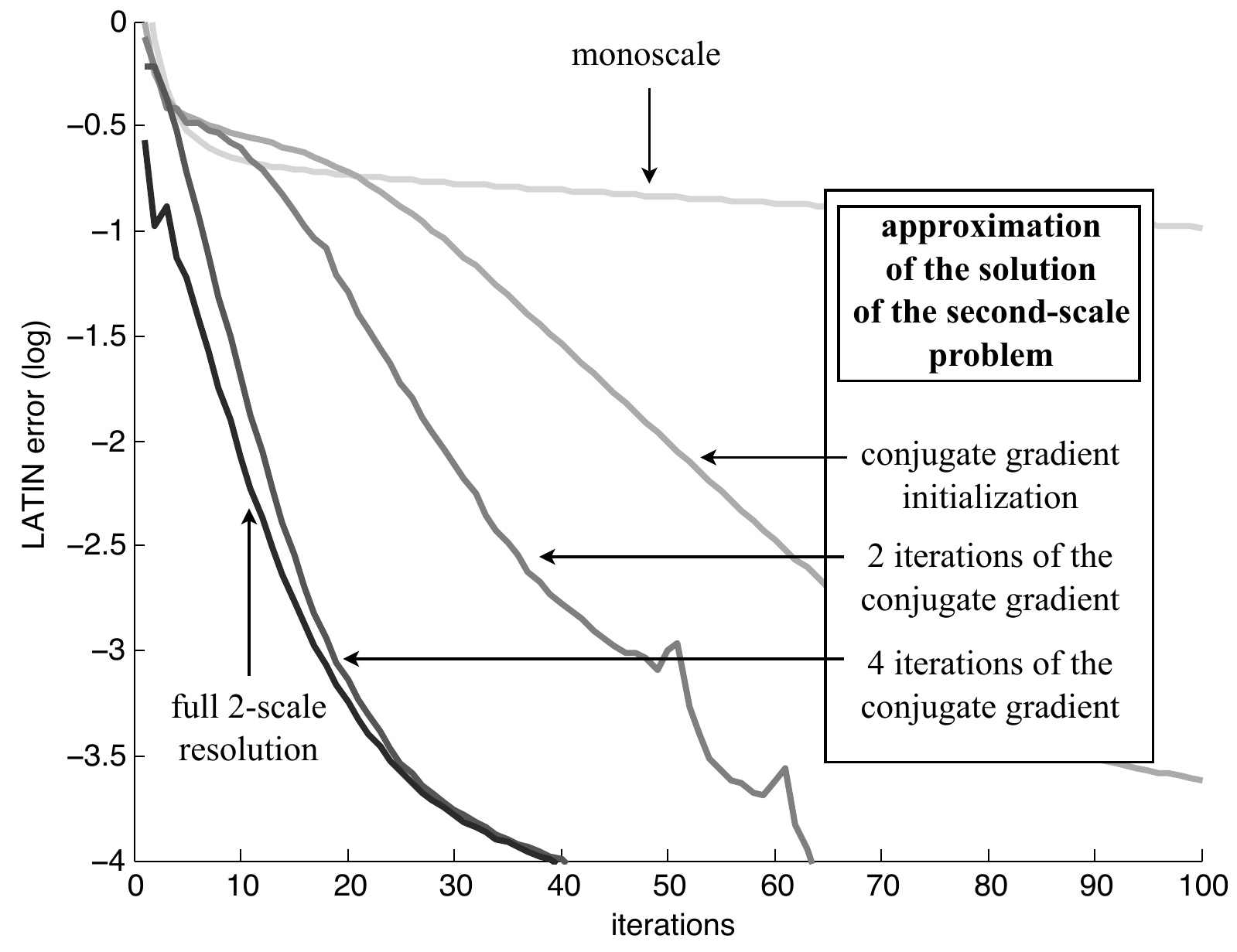}
       \caption{LaTIn convergence curves for different numbers of macro iterations}
       \label{fig:cg_conv}
\end{figure}

Matrix $(\mathbf{C}^T \mathbf{S} \mathbf{C})$ corresponds to a
coarse representation of the global stiffness of the structure.
Operator $\mathbf{C}$ must contain at least the rigid body modes of
the super-substructure. Then, the initialization ${\V{\widetilde{W}}_b}_0$ is achieved
as a combination of the rigid body modes of the local stiffness
operators ($\operatorname{range}( \mathbf{C} )$), while the
remaining part is sought iteratively in the supplementary subspace
$\operatorname{ker}( \mathbf{C}^T \mathbf{S})$ through the projector
$\mathbf{P}$.

\begin{figure}[h]
       \centering
       \includegraphics[width=0.99 \linewidth]{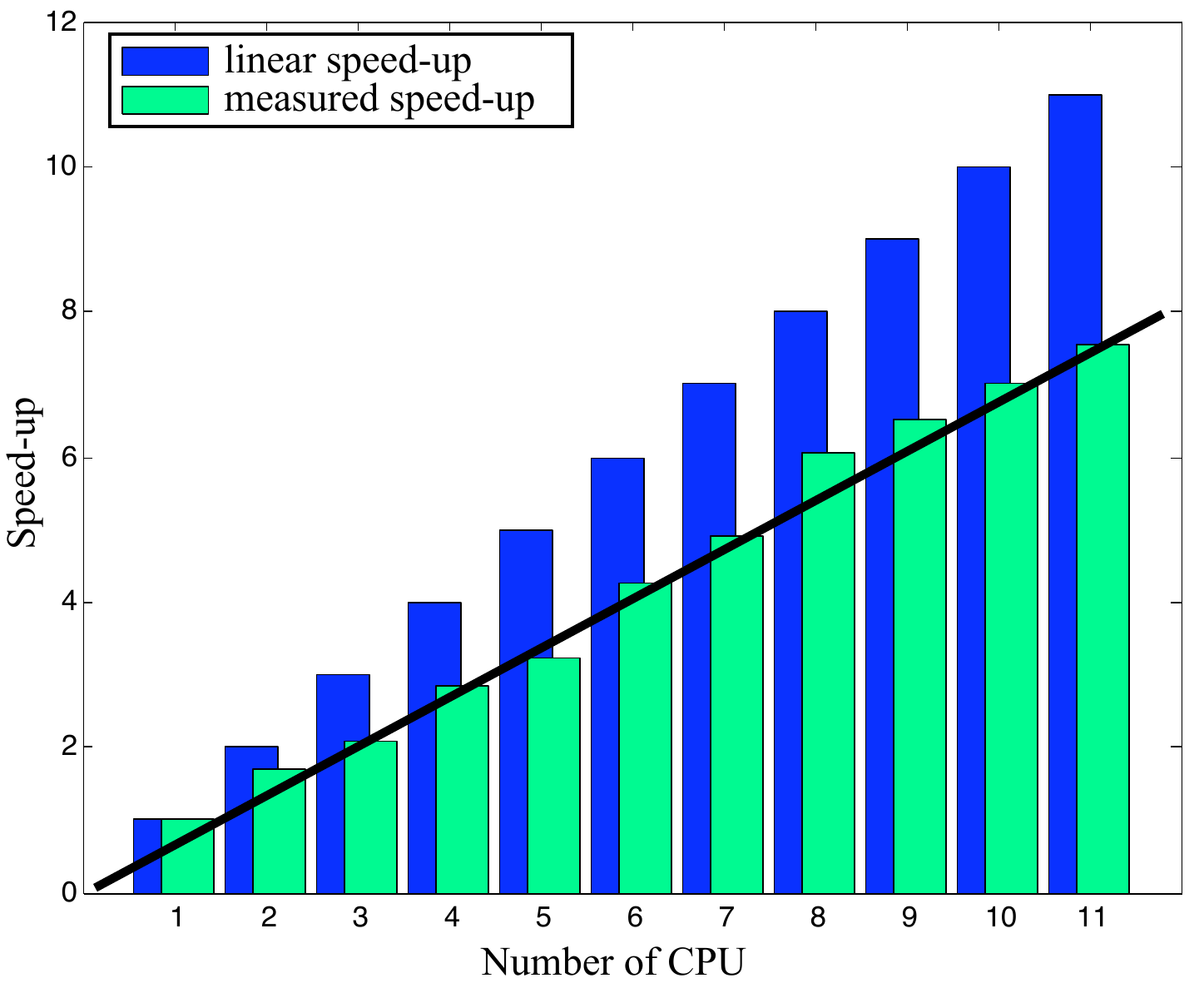}
       \caption{LaTIn convergence curves for different numbers of macro iterations}
       \label{fig:speed_up}
\end{figure}

\begin{figure}[h]
       \centering
       \includegraphics[width=0.99 \linewidth]{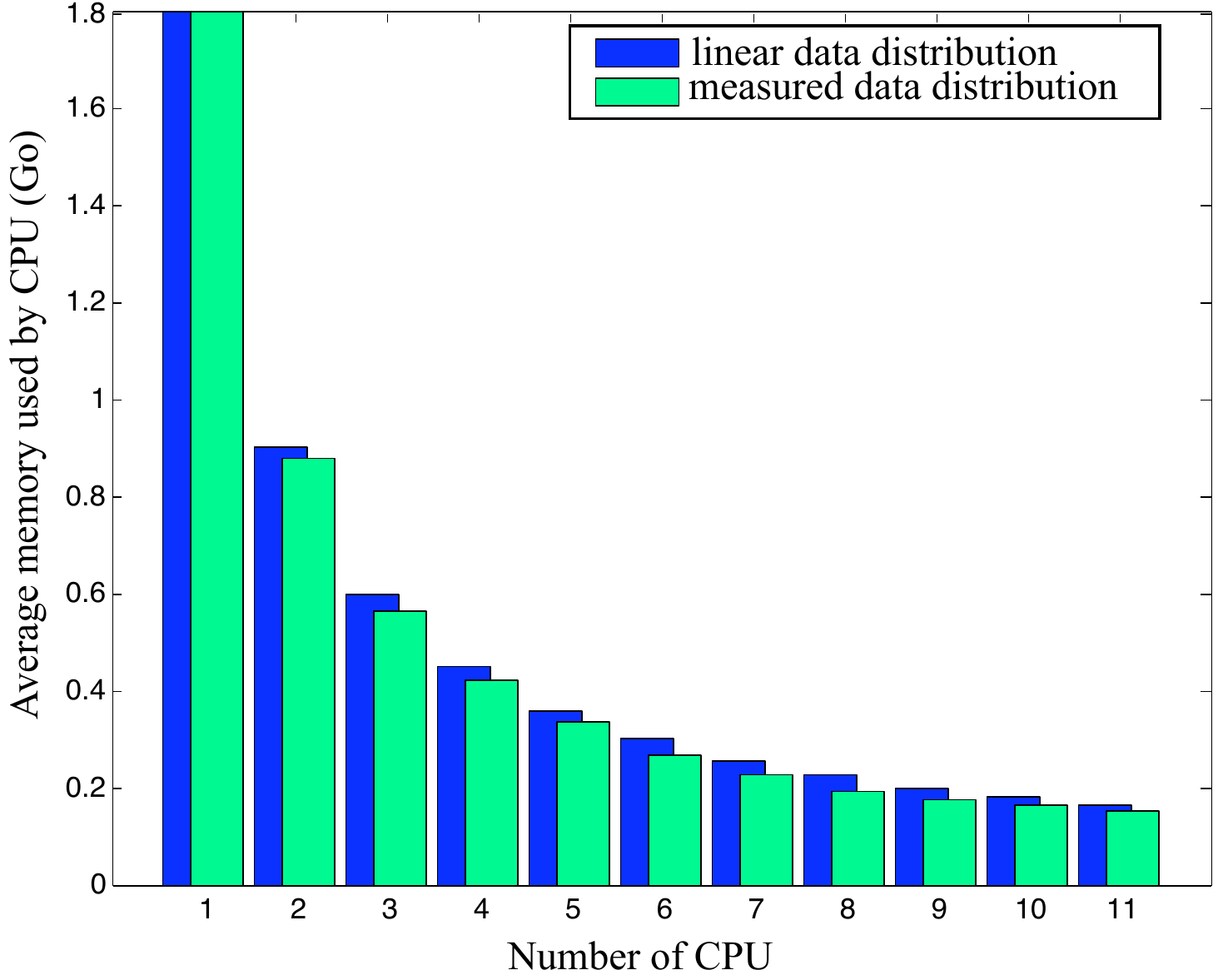}
       \caption{LaTIn convergence curves for different numbers of macro iterations}
       \label{fig:memory}
\end{figure}

\subsection{Results}

Figure (\ref{fig:cg_conv}) shows the convergence rate of the LaTIn
algorithm when the conjugate gradient scheme for the condensed
macroproblem is stopped after a fixed number of iterations. The test
case is the holed plate under traction loading represented in
Figure (\ref{fig:holed}) with the super-substructuring pattern given
in Figure (\ref{fig:third_scale}). The structure is divided into 520 substructures, separated by 1350 interfaces. The number of micro (respectively macro) degrees of freedom involved in this problem is $3.4 \times 10^6$ (respectively 12150).
It appears clearly that only very
few iterations of the conjugate gradient scheme are required to get
the necessary part of the macrodisplacement Lagrange multiplier
leading to the multiscale convergence rate of the LaTIn algorithm. Hence, high accuracy of the macroscopic resolution is not necessary to transmit pertinent piece of information on the whole structure. Typically, the algorithm is stopped when the residual error
(normalized by the initial error) falls below $10^{-1}$. 
The admissibility of the macroforces is thus enforced on a third level,
which is sufficient to determine the part of the solution which
needs to be transmitted at each iteration of the resolution.

Figures (\ref{fig:speed_up}) and (\ref{fig:memory}) show how well the method scales both in computational time  and memory usage when employed on modern hardware parallel architectures (distributed memory clusters).

\section{Control of the loading sequence}
\label{sec:riks}
The incremental version of the LaTIn algorithm, like Newton-Raphson algorithm encounters numerical difficulties when used to carry on analysis beyond global limit-points or snap-backs.


This section focuses on the discretized assembled nonlinear problem :
\begin{equation}
\mathbf{K} \left((U_m)_{m \leq n} \right) \, U_n = F_n 
\end{equation} 
Global equilibrium states are sought successively at each time step $n $ $(0 < n \leq N)$ using an implicit time integration scheme. The nonlinear equilibrium problems coupled with a local arc-length control are solved by a modified Newton algorithm, the linear prediction steps being handled by the 3-scale domain decomposition strategy, as described in next subsection.

The control algorithm is activated locally during the time analysis when the LaTIn strategy fails to converge. Conversely when the control algorithm results in a succession of re-increasing loadings, the solution algorithm is switched back to the LaTIn method.

\subsection{Local control}

As usual in arc-length methods \cite{riks72,crisfield81}, the amplitude of the loading
$\lambda_n$ is linked to the global displacement in such a way that
the norm of the $(\Delta U_n , \Delta \lambda_n \, F)$ takes a
predefined fixed value, the $\Delta \, . \,$ unknowns being the increments
of these quantities between time steps $n-1$ and $n$. In our case,
these global unknowns are not of primary interest \cite{schellekens93,allix96}. Instead, let us
introduce the control equation as:
\begin{equation}
\mathbf{c}(U_n) \, \Delta U_n = \Delta l
\end{equation}
where $\mathbf{c}(U)$ is a Boolean operator extracting the maximum
value of the local displacement gap over all the cohesive interfaces
of the model, and $\Delta l$ is a given value. Thus, the loading
increment $\lambda$ is controlled by a local variable which is
closely related to the maximal local damage increment of the structure.

\subsection{The arc-length resolution}

Thus, at each time step $n$, we seek the solution $(U_n,\lambda_n)$
of the following discrete system:
\begin{equation}
\left\{ \begin{array}{l}
\displaystyle f(U_n,\lambda_n) = \mathbf{K}(U_n) \, U_n - \lambda_n \, F = 0 \\
\displaystyle g(U_n,\lambda_n) = \mathbf{c} (U_n) \, \Delta U_n - \Delta l= 0
\end{array} \right.
\label{eq:riks}
\end{equation}
The first-order expansion of the equilibrium equation around the
point $(U_{i},\lambda_{i})$ yields:
\begin{equation}
\left\{ \begin{array}{l}
\begin{array}{ll}
\displaystyle f(U_n^{i+1},\lambda_n^{i+1}) = & \displaystyle f(U_n^{i},\lambda_n^{i}) + \frac{\partial f}{\partial U}_{|(U_n^i,\lambda_n^i)} \ (U_n^{i+1}-U_n^i) \\ 
& \displaystyle + \frac{\partial f}{\partial \lambda}_{|(U_n^i,\lambda_n^i)} \ (\lambda_n^{i+1} - \lambda_n^{i}) = 0
\end{array}
 \\
 \begin{array}{ll}
\displaystyle g(U_n^{i+1},\lambda_n^{i+1}) = & \displaystyle g(U_n^{i},\lambda_n^{i}) + \frac{\partial g}{\partial U}_{|(U_n^i,\lambda_n^i)} \ (U_n^{i+1}-U_n^i)  \\
& +  \displaystyle\frac{\partial g}{\partial \lambda}_{|(U_n^i,\lambda_n^i)} \ (\lambda_n^{i+1} - \lambda_n^{i}) = 0
\end{array}
\end{array} \right.
\end{equation}

We use a modified Newton algorithm, which means that the tangent
operator $\frac{\partial f}{\partial U}_{|(U_n^^,\lambda_n^i)}$ is
approximated by $\mathbf{K}(U_n^i)$, while $ \frac{\partial
g}{\partial U}_{|(U_n^i,\lambda_n^i)}$ is approximated by $\mathbf{c}
(U_n^i)$. Using the relations expressing that System \eqref{eq:riks}
is verified at time step ($n-1$), one gets:
\begin{equation}
\left\{ \begin{array}{l}
\displaystyle \Delta U_n^{i+1} = \lambda_n^{i+1} \, \mathbf{K}(U_n^i)^{-1} F - U_{n-1} \\
\displaystyle \mathbf{c}(U_n^i) \, \Delta U_n^{i+1}  = \Delta l
\end{array} \right.
\end{equation}
The introduction of the linearized equilibrium into the linearized
control equation leads to the expression of the loading parameter at
Iteration ($i+1$), then to the displacement solution:
\begin{equation}
\left\{ \begin{array}{l}
\displaystyle \lambda_n^{i+1} = \frac{\Delta l + \mathbf{c}(U_n^i) \, U_{n-1}}{\mathbf{c}(U_n^i) \, \mathbf{K}(U_n^i)^{-1} F}\\
\displaystyle U_n^{i+1}  = \lambda_n^{i+1} \, \mathbf{K}(U_n^i)^{-1} F
\end{array} \right.
\label{eq:predic_riks}
\end{equation}

Then, Operators $\mathbf{K}(U_n)$ and $\mathbf{c}(U_n)$ are updated with respect to the kinematic field $U_{i+1}$ found at the prediction stage of the modified Newton algorithm \eqref{eq:predic_riks} and, unless the residuals of the updated equilibrium and control equation are small enough, a new iteration is performed. The norm of the residual of the control equation is not used to stop the Newton iterations. In facts, the given value of the maximum local damage increment has no physical meaning. Yet, using this numerical technique ensures that the evolution of the loading permits to follow the global behavior of the structure. Consequently, any converged equilibrium state found can be used to perform a new time-step computation.

\subsection{Parallel calculation}

Our attempt to solve the linear problem \eqref{eq:riks} has been unsuccessful. This can be explained by the fact that the control equation is global over the whole structure and non-linear. Thus the classical separation of the linear equations on one hand and the local non-linear equations on the other hand, which is the basic idea of the LaTIn method, cannot be made. Nevertheless, the prediction step of the Newton algorithm described in the previous section requires the resolution of the linear system $\mathbf{K}(U_n^i)^{-1} F$. We propose to use the LaTIn mixed three-scale domain decomposition method  to find the solution to this linear system. The method is still efficient for two reasons :
\begin{itemize}
\item The non-linearities being computed are still localized on the interfaces of the domain decomposition method. Consequently, no re-assembling step is required as the Newton iterations proceed.
\item using a full nonlinear LaTIn solver or using a Riks solver with a parallel LaTIn resolution of the prediction step requires very similar computations, which means that switching from the LaTIn algorithm to the arc-length algorithm is straightforward
\end{itemize}

Nevertheless, one should notice that the linearity of the prediction step of the Riks solver makes the subiteration technique described in Section (\ref{sec:subiterations}) non-relevant (or at best less effective).

\begin{figure}
       \centering
       \includegraphics[width=0.99 \linewidth]{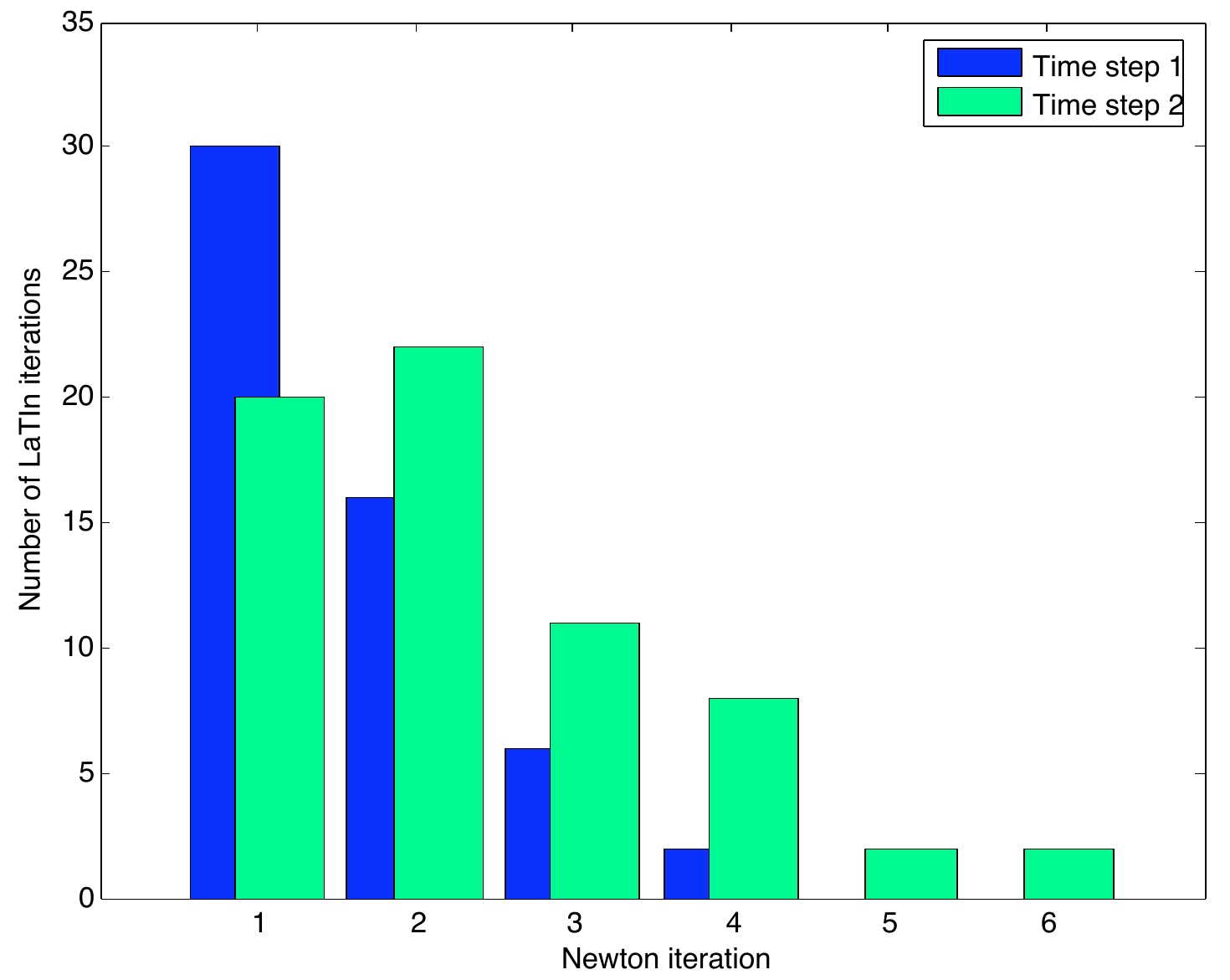}
       \caption{Influence of the initialization of the LaTIn linear solver on the number of LaTIn iterations required to achieve convergence at each Newton iteration}
       \label{fig:multires}
\end{figure}

\begin{figure}
       \centering
       \includegraphics[width=0.99 \linewidth]{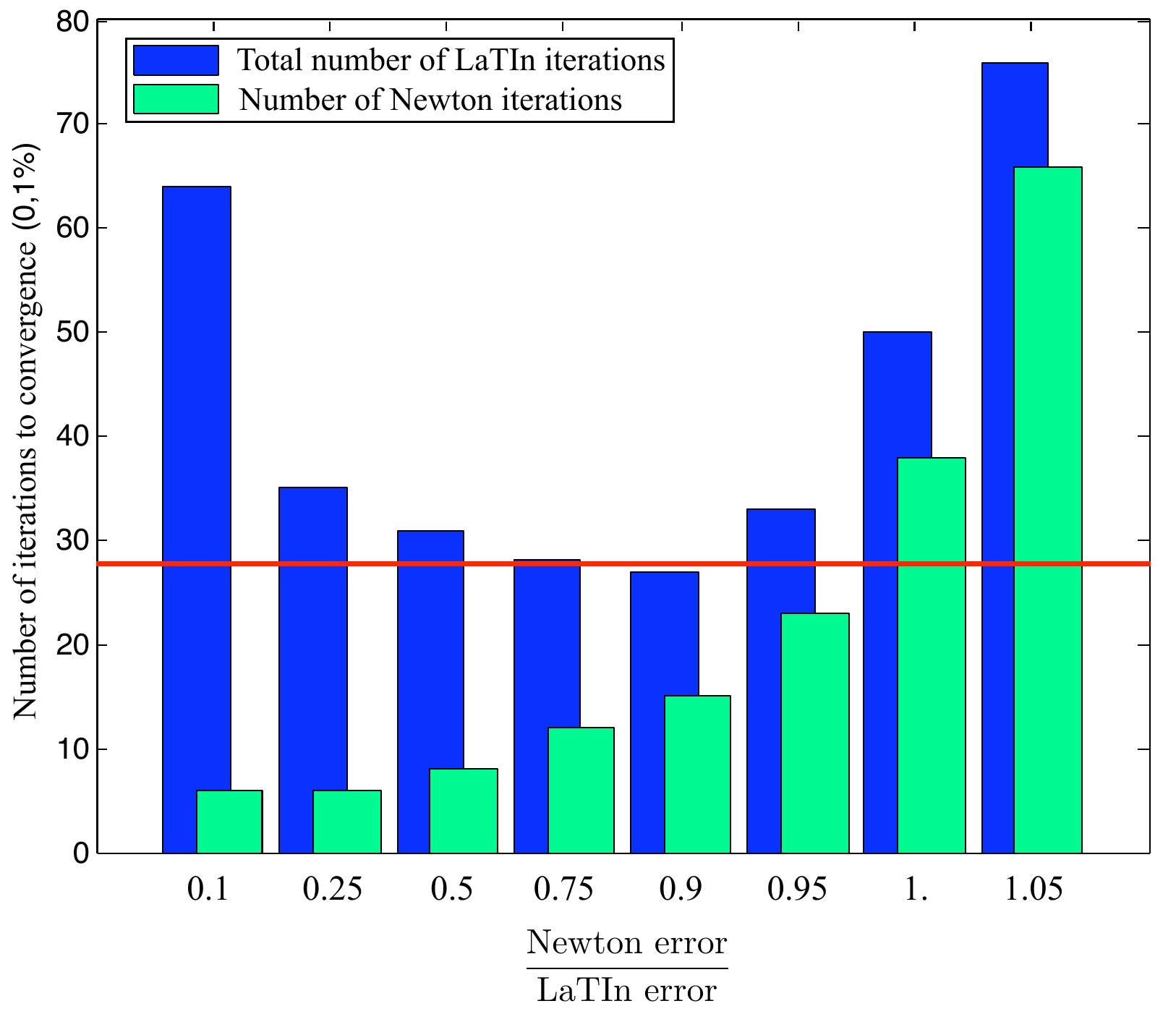}
       \caption{Influence of the stopping criteria on the numbers of iterations of the Newton algorithm and of the LaTIn linear solver, to compute one Newton increment}
       \label{fig:nr_latin}
\end{figure}

\begin{figure*}
       \centering
       \includegraphics[width=0.7 \linewidth]{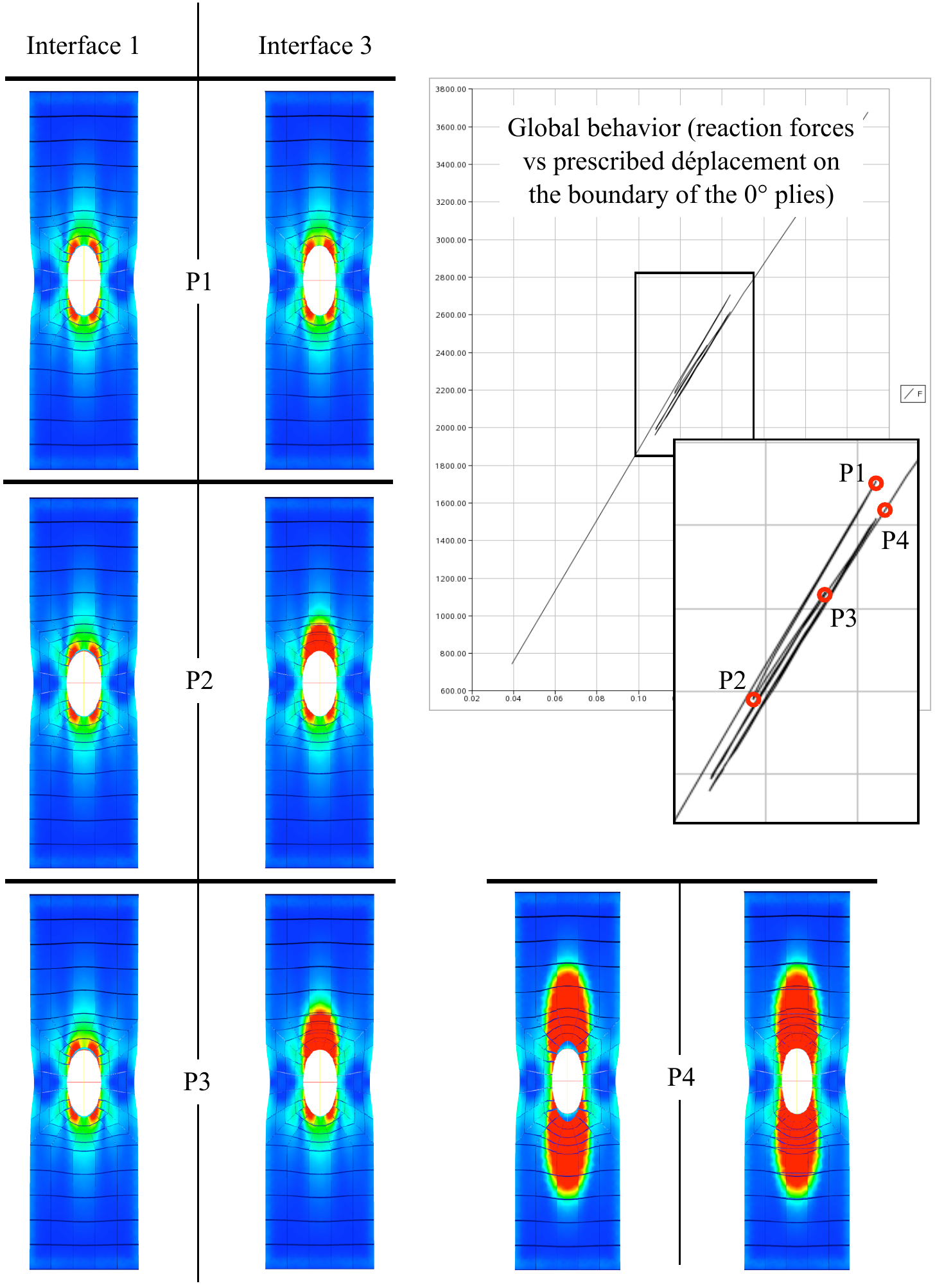}
       \caption{Behavior of the holed plate under traction}
       \label{fig:riks}
\end{figure*}

\subsection{Numerical improvement of the Newton algorithm using the LaTIn method as a linear solver}

Two elements can reduce the cost of the arc-length algorithm significantly:
\begin{itemize}
\item the initialization of each iteration of the local arc-control algorithm with the interface quantities found after convergence of the LaTIn resolution scheme in the previous iteration. The number of LaTIn iterations decreases very rapidly as the iterations of the Newton algorithm proceed because the changes in the secant operators $\mathbf{c}(U_n)$ and $\mathbf{K}(U_n)$ become less and less significant. On Figure (\ref{fig:multires}), the two first time steps of the holed plate test case (see Figure (\ref{fig:holed})), which correspond to different levels of non-linearity, are computed using a stopping criteria of the LaTIn linear solver set to a given value. The drop of the number of LaTIn iterations required as the Newton algorithm goes on appears clearly in these two cases. As this idea systematically improves convergence for no extra cost, it is now used by default.
\item the crossed-optimization of the stopping criteria of (non-linear) Newton solver and (linear) LaTIn solver.
Basically, the LaTIn algorithm could be converged to a very low value of the residual of the linear system at each prediction step of the Newton scheme. Though, our tests show that the first iterations of Newton, leading to a high value of the residual of the non-linear system, do not require an exact resolution of the prediction step. In order to illustrate this idea,
the number of Newton iterations for one time step of a DCB test (Figure (\ref{fig:adap_endom} top)), and the associated total number of (linear) LaTIn iterations are plotted in Figure (\ref{fig:nr_latin}) as functions of the ratio of these two errors . This shows clearly that in order to use the method most efficiently (\textit{i.e.} with the smallest total number of LaTIn iterations), the convergence threshold of the LaTIn method should be set to an error very close to the current Newton error.
\end{itemize}

\subsection{Results}

Figure (\ref{fig:riks}) shows the global force \textit{vs.} displacement curve obtained for the holed plate test
case (\ref{fig:holed}) using the arc-length algorithm described above. The damage in the interfaces loaded in shear mode (interfaces [0/90]) is also represented at four equilibirum states of the time analysis. Several very sharp snap-backs appear in the global behavior curve of the structure, and are efficiently handled by this locally controlled Riks' algorithm.

\section{Conclusion}

The accurate prediction of delamination in large process zones of laminate composite structures requires refined models of the material behavior. Such descriptions lead to the resolution of huge systems of equations. In order to calculate the exact solution of such a refined model, we used a two-scale domain decomposition strategy based on an iterative resolution algorithm. This method is particularly well-suited for laminate models in which 3D and 2D entities are introduced separately.

This strategy has been improved in order to make it capable of handling very large delamination problems. A systematic analysis of the features of the method at the different scales has been conducted. It has first been shown that the classical scale separation was insufficient in the high gradient zones to provide numerical scalability. We thus developed a subresolution procedure which preserved the numerical scalability of the crack propagation calculation. This analysis has also proved that a third scale was required. The second-scale problem is then solved using a parallel iterative algorithm, which enabled the fast transmission of the very-large-wavelength part of the solution.

In order to perform the quasi-static analysis beyond the global limit-points resulting from the local softening behavior of the structure, we used an arc-length-type algorithm to control the magnitude of the loading. We showed that the computation steps required when using this algorithm were very similar to those of the LaTIn technique. Therefore, switching from one algorithm to the other was very easy. 

In future developments, this 3D process zone analysis technique should be associated with a plate model analysis, which would be sufficient to obtain the solution in the low-gradient zones.

%
%

\bibliographystyle{plain}
\bibliography{template}

%
%

\end{document}